\documentclass[12pt,preprint]{aastex}

\shorttitle{Multipole Accretion Funnel}
\shortauthors{Mohanty \& Shu}

\begin{document}

\def\zw{z_w }
\def\dop{\mathcal{D} }
\def\xangle{\vartheta }
\def\xanglew{\vartheta_w }
\def\Phid{{\Phi}_d }
\def\Phidx{{\Phi}_{\rm dx} }
\def\Phimx{{\Phi}_{\rm mx} }
\def\Phim{{\Phi}_{\rm m}}
\def\Phif{{\Phi}_{\rm f} }
\def\Phit{{\Phi}_t }
\def\Phiw{{\Phi}_w }
\def\Phir{{\Phi}_{\rm r} }
\def\rx{R_X }
\def\rk{R_k }
\def\del{{\bf{\nabla}} }
\def\delsq{{\nabla}^2 }
\def\vecr{{\bf{r}} }
\def\vecB{{\bf{B}} }
\def\vecBp{{\bf{B}_{\rm p}} }
\def\vecBa{{{B_{\varphi}}} }
\def\vecBr{{{B_{\varpi}}} }
\def\vecBz{{{B}_z} }
\def\malven{{\mathcal{M}}_A }
\def\rstar{R_{\ast} }
\def\rd{R_D }
\def\omstar{{\Omega}_{\ast} }
\def\omx{{\Omega}_X }
\def\ax{a_X }
\def\jbstar{\bar J_{\ast} }
\def\zmax{z_{\rm max} }
\def\deg{$^{\circ}$ }
\def\erhat{{\^{e}}$_{\varpi}$}
\def\eahat{{\^{e}}$_{\varphi}$}
\def\ezhat{{\^{e}}$_z$}
\def\msun{M$_{\odot}$ }

\title{Magnetocentrifugally Driven Flows from Young Stars and Disks.\\ 
VI. Accretion with a Multipole Stellar Field}
\author{Subhanjoy Mohanty\altaffilmark{1} \& Frank H. Shu\altaffilmark{2}}
\altaffiltext{1}{Harvard-Smithsonian CfA, 60 Garden Street, Cambridge, MA
02138, U.S.A.  smohanty@cfa.harvard.edu}
\altaffiltext{2}{Physics Department, University of California at San Diego, La Jolla, CA 92093, USA}

\begin{abstract}
Previous analyses of magnetospheric accretion and outflow in classical T Tauri stars (CTTSs), within the context of both the X-wind model and other theoretical scenarios, have assumed a dipolar geometry for the stellar magnetic field if it were not perturbed by the presence of an accreting, electrically conducting disk.  However, CTTS surveys reveal that accretion hot spots cover a small fraction of the stellar surface, and that the net field polarization on the stellar surface is small.  Both facts imply that the magnetic field generated by the star has a complex non-dipolar structure.  To address this discrepancy between theory and observations, we re-examine X-wind theory without the dipole constraint. Using simple physical arguments based on the concept of trapped flux, we show that a dipole configuration is in fact not essential.  Independent of the precise geometry of the stellar magnetosphere, the requirement for a certain level of trapped flux predicts a definite relationship among various CTTS observables.   Moreover, superposition of multipole stellar fields naturally yield small observed hot-spot covering fractions and small net surface polarizations.  The generalized X-wind picture remains viable under these conditions, with the outflow from a small annulus near the inner disk edge little affected by the modified geometry, but with inflow highly dependent on the details of how the emergent stellar flux is linked and trapped by the inner disk regions.  Our model is consistent with data, including recent spectropolarimetric measurements of the hot spot sizes and field strengths in V2129 Oph and BP Tau. 

\end{abstract}

\keywords{stars: low mass -- stars: pre-main-sequence --circumstellar matter -- stars: formation}

\section{Introduction}
Various theoretical models have been proposed for the physical mechanisms driving the accretion and outflow processes
in classical T Tauri stars (CTTSs), with the most popular being perhaps Blandford and Payne's (1982) pioneering study of magnetocentrifugally driven winds from Keplerian disks. However, the X-wind model has gained credence in recent years for a variety of theoretical and observational reasons (see also the discussion in \S 5).  It is instructive at the outset to summarize and compare the different CTTS models, to review the rationale for X-wind theory and the evidence in its favor, and to motivate our subsequent generalization of this picture (see also Shu et al. 2000).  

The first accretion models of CTTSs proposed that they were surrounded by Keplerian disks, which extend all the way to the stellar surface, with accretion occuring through an equatorial boundary layer (e.g., Bertout 1987; Kenyon \& Hartmann 1987; Bertout et al. 1988).   Motivated by the observational finding that the driver behind some well-known bipolar outflow sources are neutral winds containing H I and CO (Lizano et al. 1988), a combination not seen in the interstellar medium but present in the photospheres of cool stars, Shu et al. (1988) suggested that the driving outflows are caused by boundary-layer disk-accretion onto a strongly magnetized young stellar object (YSO).  The accretion provides a rationale for why a protostar might rotate near break-up, and Hartmann \& MacGregor (1982) had already shown that magnetized stars rotating near break-up could shed matter and angular momentum extremely efficiently through their equatorial zones.  However, Shu et al. (1988) obtained preliminary indications that streamline collimation in this kind of model was absent, or extremely slow, and therefore they speculated stellar jets were associated with ``ordinary'' stellar winds confined to flow along the rotation axis by the more powerful ``extraordinary'' centrifugally-driven outflow that they later called an X-wind.  

When a disk abuts a fully convective star, Ekman pumping (internal circulation caused by slight pressure differences in
a field of differential rotation) to the equatorial boundary layer adjoining the star and the disk eventually causes the entire star to spin near break-up if there are no countervailing spin-down torques (Galli 1990; see Fig. 10 of Shu et al. 1993).  CTTSs are fully convective, yet they usually rotate at velocities one order of magnitude below break-up (e.g., Vogel \& Kuhi 1981; Hartmann \& Stauffer 1989; Bouvier et al. 1993, 2007).  Shu et al. (1988) therefore speculated that CTTSs are young stars in which the spin-down by ordinary stellar winds overcame the spin-up toques of the viscous boundary-layer, and that CTTSs would not have strongly collimated outflows.  

The latter expectation turned out to be false (see, e.g., Edwards et al. 1993).  Nevertheless, Matt \& Pudritz (2005, 2008a, b)
have recently resurrected the idea that magnetized stellar winds might provide sufficient torque to explain the slow rotation
of T Tauri stars and, perhaps, be responsible for a part of the observed optical jets from YSOs.  The latter proposal using coronal gas to launch the stellar winds has, however, foundered on the constraints provided by X-ray observations of T Tauri stars (Bisnovatyi-Kogan \& Lamzin 1977, DeCampli 1981, Matt \& Pudritz 2008c), and the finding that CTTS stellar winds appear to be an order of magnitude cooler (Johns-Krull \& Herczeg 2007) than suggested recently (Dupree et al. 2005).  
The issues, therefore, became {\it (1)} how to accrete high angular momentum disk material onto a CTTS while keeping the latter rotating slowly, {\it (2)} how simultaneously to generate winds efficiently from the star+disk system, and {\it (3)} how to make the outflowing wind appear as a jet despite the slow streamline collimation? 

K\"{o}nigl (1991) suggested a solution for the accretion part of this problem by adopting the theory developed by Ghosh \& Lamb (1978; 1979a, b).  A strong stellar magnetosphere is assumed to truncate the disk some distance from the stellar surface, with accreting material flowing onto the star not through a boundary layer but via magnetic field lines threading the disk.  The angular momentum of the star is then regulated by the interaction between the stellar magnetosphere and the disk.  Field lines originating in the star and with disk footpoints within the Keplerian corotation radius ($\rx$) are dragged forward by the disk gas, relative to the star, and thus tend to spin the star up (at the expense of the disk gas).  Stellar field lines threading the slower rotating disk beyond $\rx$, on the other hand, tend to spin the star down.  The slow rotation of CTTS then arise if the spin-down caused by the outer disk outweighs the spin-up by the inner parts. Large turbulent diffusivities must prevail to allow magnetic fields to slip through the gas in such a picture.

In a series of papers with a precedent in the suggestions of Arons (1986) and Camenzind (1990), Shu and collaborators (Shu et al. 1994a, b; Najita \& Shu 1994; Ostriker \& Shu 1995, hereafter OS95; Shu et al. 1995) embraced the idea that strong YSO magnetospheres might truncate the disk before it abuts the stellar surface, but they pointed out a number of difficulties with the specific proposal of K\"onigl (1991).  The resolution of these difficulties turned out to provide solutions for each of the issues {\it (1)} to {\it (3)} listed above.

First, the picture painted by Ghosh \& Lamb and K\"onigl changes considerably if turbulent resistivities are not large, but small, i.e., if field diffusion is competitive with advective flow not on dynamical time scales but secular ones.  In such cases, the long-term processes of angular-monetum transport outwards and mass-transport inwards dominate over notions of ``ram-pressure balance''.  Second, stellar fields strong enough to truncate an electrically conducting disk are automatically also strong enough to drive a magnetocentrifugal outflow along the outermost flux tubes of the stellar field, which are opened into an X-wind.  The magnetic torques in the wind cause the outflowing matter to gain angular momentum at the expense of the material still connected to it by field lines threading through the disk.  The back reactions to the X-wind and funnel-flow give a pinch of the exterior field lines inward and the interior field lines outward towards a common mid-point $\rx$, {\it with a net trapped flux}, that gives the X-wind model its name.
This mid-point is both in Keplerian rotation and in corotation with the star, i.e., $\Omega(R_X) = (GM_*/R_X^3)^{1/2} = \Omega_*$, a condition that came to be called {\it disk locking}. 
Third, isodensity contours become cylindrically stratified very quickly after the gas accelerates from the X-region (Najita \& Shu 1994; Shu et al. 1995), yielding the optical illusion in the emission of forbidden-line and radio emission that X-winds achieve jet-like collimation close to the base of the flow (Shang et al. 2002, 2004).  In actual practice, streamline collimation is logarithmically slow, which has observable consequences for position-velocity diagrams obtained from long-slit spectrograms (e.g., Pyo et al. 2006).

Many numerical simulations have also been devoted to the X-wind/funnel flow problem, the most successful being that of Romanova et al. (2008).  Indeed, the progress made by the simulations is most concisely revealed by examining why they succeeded in obtaining the simultaneous existence of X-winds and funnel flows over extended durations when others failed.  As stated in the previous paragraph and detailed in \S 2, the crucial concept in X-wind/funnel flow theory is that of the {\it trapped flux} created by the two-sided pinching of field lines toward $\rx$.  In the presence of non-zero resistivity $\eta$, field diffusion will occur out of the X-region.  This diffusion must be offset by fluid advection from both the exterior and the interior of $\rx$.  

In the exterior, disk inflow is induced both from back-reaction from the X-wind (if present) and by viscous inflow from the disk proper, with the level of kinematic viscosity $\nu$ dictating the disk accretion-rate $\dot M_D$.
If the viscosity $\nu$ is only comparable to the resistivity $\eta$, the disk inflow is too weak relative to diffusive penetration to produce a good inward pinch, a failure not conducive to the generation of X-winds.  To produce sufficiently outward-bending field lines from viscous accretion, it is necessary to have $\eta$ smaller than $\nu$ (by a factor of roughly the disk-aspect ratio $z_0/\varpi \ll 1$; see Lubow, Papalozou, \& Pringle 1994 and Shu et al. 2007).  For the specific problem of an accretion disk interacting with a stellar magnetosphere, the simulations of Romanova et al. (2008) demonstrate explicitly that the condition $\eta \ll \nu$ is indeed the crucial ingredient to achieving an X-type magnetic configuration and thereby generating an X-wind.

In the interior, a funnel flow involving closed field lines links the star to the disk.  If the stellar rotation rate $\Omega_*$ is not chosen to be the same as the angular speed $\sim \Omega_X$ of the parts of the disk to which the filed lines are rooted, rapid transients are induced.  Some early simulations, which started with slowly rotating stellar magnetospheres linked to rapidily rotating disks in the initial state, managed to obtain temporary X-type magnetic configurations via such transients (Hayashi et al. 1996, Miller \& Stone 1997, K\"uker et al. 2003).  The sudden removal of angular momentum from the disk by magnetic torques creates a dynamic onrush of material toward the star, generating a temporary magnetic pinch and outflow.  In these studies, finite resistivity is sometimes included in the code, but this resistivity plays no important role because the magnetic diffusion is slow compared to the fast inward flow caused by the large disequilibrium of the initial state.

Goodson et al. (1997, 1999), Goodson \& Winglee (1999), and Romanova et al. (2002, 2003, 2004a, 2005, 2008; hereafter R02, R03, R04, R05, R08) pioneered the incorporation of more realistically rotating stellar magnetospheres, with the later papers of the latter group and Goodson et al. (1997, 1999) using controlled levels of resistivity and viscosity. In particular, R02 and R03 performed axisymmetric and non-axisymmetric calculations that start with an unperturbed, aligned and tilted, stellar dipole field threading a circumstellar disk that begins to accrete slowly via a postulated disk viscosity $\nu$.  The only resistivity in the problem is numerical and gives an effective value for  $\eta$ comparable to $\nu$.  Although no X-winds arose as a consequence, these authors did show that in steady-state the radius $R_{\rm co}$ where the magnetosphere corotates with the disk is close to the disk truncation radius (denoted by them as the stellar magnetopause $R_{\rm m}$), with the best runs having $R_{\rm co}/R_{\rm m}$ = 1.2--1.3 (see also Long et al. 2005).  Using the same basic configuration, but rotating the star more quickly so that corotation is reached interior to the truncation radius -- the so-called ``propeller regime'' (Illarionov \& Sunyaev 1975) -- R05 found strong disk outflows to be possible if $\nu$ is several times larger than $\eta$ (see also Ustyugova et al. 2006; hereafter U06).  The simulations of R04 discovered ``magnetic towers'' in the stellar corona near the rotation axis as first proposed by Draine (1983) and Lynden-Bell (1996).  With realistic levels of coronal density, such towers do not carry much matter and cannot explain YSO jets (Long et al. 2005; see also Figs. 4 \& 5 of Allen et al. 2003b for the appearance of the "magnetic tower" phenomenon in the context of the collapse of a rotating, magnetized, molecular-cloud core).  Of specific interest to our present paper, the lowering of the coronal densities assumed by R04 by two orders of magnitude in the simulations of Long et al. (2005) made much easier the opening of stellar field lines, and led the way to the driving of powerful YSO jets in an X-type magnetic configuration by R08 when $\eta \ll \nu$ and $R_{\rm co} \approx R_{\rm m} \approx R_X$. 

The main remaining differences between ideal X-wind theory and numerical simulations concerns whether outflows can be steady.  Figure 1 shows that field lines dead to inflow or outflow (black) separate the funnel flow (red) from the X-wind (blue).  Some of the dead-zone field lines are closed and link the disk to the star; some are open and ``joined at infinity'' line by line to the magnetic field contained in the X-wind.  The reversal of field direction across the separatrix between the open dead-zone fields and the open X-wind fields of ideal MHD would be unstable to reconnection events in the presence of finite resistivity.  These reconnection events, which are likely to be episodic (Aly \& Kuijpers 1990), would create a fluctuating X-wind (Shu et al. 1997; Romonova et al. 1998; Uzdensky, K\"{o}nigl, \& Litwin 2002), and may underlie the outbursting behavior found in the simulations of Goodson et al. (1997, 1999), R04, R05, U06, and R08.  The steady assumption of ideal X-wind theory is then made for analytic simplicity, and can at best represent only the time-average of outflows that are time-variable and quasi-periodic in reality.

\section{Generalized X-Wind Model}

To date, most of the formal developments of X-wind theory surmise that the magnetic field configuration of the star in the absence of any interactions with the disk is a pure dipole.  But this is a dubious simplification because there is no reason to expect that the dynamo action in slowly rotating, fully convective objects (such as T Tauri stars) will create a highly organized pattern of surface magnetic fields.  Indeed, polarization studies indicate that an organized dipole component cannot be dominant on the surfaces of most T Tauri stars (Valenti \& Johns-Krull 2004).  Moreover, the observed covering fraction of ``hot spots'' due to funnel flows is typically much smaller (0.1 - 1\% of the stellar surface) than the predictions from dipole models (Johns-Krull \& Gafford 2002, hereafter JG02, and references therein; see also Gregory et al. 2006a and Jardine et al. 2008).  Recent numerical simulations by Chabrier \& K\"{u}ker (2006) similarly predict that in fully convective objects, the ${\alpha}^2$ dynamo effect should produce large-scale non-dipolar magnetic fields, dominated by higher-order multipoles.  

Numerical simulations have kept pace with the observational developments.  Thus, Romanova et al.~(2004b) have performed 3-D MHD simulations of the funnel flows that result from inclined stellar dipoles, and Long et al. (2007, 2008) have included the further superposition of quadrupole contributions to such 3-D calculations.  Relative to the observational issues raised by JG02 regarding funnel flows and their hot spots, the most important finding of these 3-D results is that the central portions of the hot spots are hotter than their peripheries, yielding the possibility that the total hot-spot covering fraction may have been under-estimated by the UV observations.

Fortunately, the general validity of the X-wind model does not depend on detailed assumptions such as the dipole approximation.  The important features are (a) the existence of trapped flux $\Phit$ in the X-region, and (b) the assumption of disk locking.  Let us see how these two ideas lead to relationships between physical parameters of the system that agree better with observations of the funnel flows of T Tauri stars than other predictions (JG02).

\subsection{Models with Trapped Flux}
We adopt cylindrical coordinates $(\varpi,\varphi,z)$ with origin at the center of the star.  Suppose that the pinch toward the X-point traps an amount of magnetic flux equal to $\Phit$ in an equatorial ring with radius $\varpi = R_X$.  In the immediate vicinity of the X-region, the trapped flux must be nearly force-free above and below the midplane (which is not force-free).  In the cold limit where the X-region corresponds essentially to the single radius $\rx$, the field configuration achieves the shape of a complete fan, and a third of all the trapped field lines bow sufficiently outward or inward to launch an X-wind or a funnel flow, carrying respectively a fraction $f$ and $(1-f)$ of the disk accretion rate $\dot M_D$.  The fraction $f$ of matter that physically climbs onto open field lines in the X-wind depends on how diffusive effects load different field lines in the X-region.  Lacking such detailed knowledge, Shu et al. proposed guessing that $f\approx 1/3$ on the basis that 1/3 of the trapped field lines in the X-region bend sufficently outward to become opened as wind streamlines. To provide the most general set of expressions admissible within our analysis, in equations (1)-(13) below we assume that the {\it flux} fractions participating in the funnel flow and X-wind are each 1/3, as imposed mathematically by the cold limit of our analysis, but, in the absence of further constraints, keep the corresponding wind {\it mass} fraction $f$ a free parameter (see further discussion at the end of \S2.2).

The field lines that are loaded with infalling gas in the funnel flow are the same ones that carry the gas all the way to the star.  Let $F_h$ be the fraction of the surface area $2\pi R_*^2$ of the upper hemisphere covered by hot spots of mean field-strength $\bar B_h$.  Then the magnetic flux connected to hot spots, $F_h (2\pi R_*^2) B_h$, must be equal to the same 1/3 of the trapped flux $\Phit$ (defined for convenience to be positive) that links to the base of the funnel flow in the upper surface of the X-region:
\begin{equation}
F_h (2\pi R_*^2) \bar B_h = {1\over 3}\Phit .
\label{hotspotflux}
\end{equation}
We will assume that the lower hemisphere looks the same as the upper hemisphere except with reversed directions of the magnetic field.  

In order to produce a self-consistent MHD flow, the trapped flux must be able to produce a time rate of change of the $z$-component of the angular momentum in the funnel flow equal to
\begin{equation}
(1-f)\dot M_D (1-\bar J_*)R_X^2\Omega_X,
\label{angmomtransport}
\end{equation}
where $R_X^2\Omega_X$ is the angular momentum per unit mass of material orbiting the star at the inner disk edge, and $\jbstar$ is the average fraction of this specific angular momentum that lands on the star (a conserved quantity carried partly by matter and partly by the magnetic torques of the field).  
An analogous argument must apply to the torque supplied to the X-wind by the flux $\Phi_t/3$ contained in it; namely, the wind torque must supply a time-rate of change of the angular-momentum equal to
\begin{equation}
f\dot M_D (\bar J_w-1)R_X^2\Omega_X .  
\label{angmomtransportwind}
\end{equation}
Thus, angular momentum enters the X-region at a rate $\dot M_D R_X^2\Omega_X$ through disk accretion, while angular momentum leaves the X-region at a rate $f\dot M_D \bar J_w R_X^2\Omega_X$ through the outflow in the wind, and at a rate $(1-f)\dot M_D \bar J_* R_X^2\Omega_X$ by accretion through the funnel flow.  In addition, there is a viscous toque ${\cal T}>0$ exerted by disk matter interior to the X-region on disk matter exterior to it. In steady state, the single term representing angular momentum entering the X-region must be balanced by the sum of the three terms leaving it.  If we solve the resulting equation for the fraction $f$ of the disk accretion rate that is carried in the X-wind, we get (see Shu et al. 1994a)
\begin{equation}
f = {1-\bar J_*- \tau\over \bar J_w - \bar J_*},
\label{f}
\end{equation}
where $\tau \equiv {\cal T}/\dot M_D R_X^2\Omega_X > 0$ is the dimensionless viscous torque acting across a circle beyond $R_X$.

Observationally, it is possible to measure $\bar J_w$ in principle from the mean terminal velocity $\bar v_w$ reached at infinity by the material in the X-wind. 
For a cold flow, conservation of energy in the corotating frame (Jacobi's constant) implies that (see Shu et al. 1994a):
\begin{equation}
\bar v_w = R_X\Omega_X \sqrt {2\bar J_w-3}.
\label{termvel}
\end{equation}
With the guesses that $\bar J_* \approx 0$ for a small, slowly rotating star, and $\tau \approx 0$, equations (\ref{f}) and (\ref{termvel}) yield $\bar J_w \approx 1/f \approx 3$ and $\bar v_w \approx \surd 3 R_X\Omega_X$, which are in rough agreement with observations of YSO winds and jets (Shang et al. 2002, 2004; Pyo et al. 2006).  To obtain the standoff distance $R_X$ for the stellar magnetopause, we need a model that gives the trapped flux $\Phi_t$ as a function of distance $R_X$ from the center of the stellar multipole distribution (see \S\S 2.2 and 4.1).
 
\subsection{Connection with Dynamics of X-Wind}

To obtain the X-point location in the meridional plane, $R_X$, it is necessary (as in OS95) to input the dynamics of the X-wind because, by assumption in the simplest model, it is the X-wind back torque that is trying to drive the inner edge of the disk inward.  The funnel flow field lines, being strong, then provide whatever (small) azimuthal component $B_\varphi$ is needed to yield a back torque that holds the inner edge in equilibrium at $R_X$. This assumption ignores any role for a magnetic tower that might torque down the spin of the star (e.g., R04, Long et al. 2005), and it is an open question in our opinion whether real YSO coronae can be sufficiently dense to make such braking important.  We assume that 1/3 of the trapped flux $\Phi_t$ drives an X-wind given by (see eq. 2.3 of OS95 and 3.10b of Shu et al.~1994b):
\begin{equation}
{1\over 3}\Phi_t = 2\pi \bar\beta f^{1/2}(GM_* \dot M_D^2 R_X^3)^{1/4},
\label{Xwind}
\end{equation}
where $\bar \beta$ is a dimensionless (inverse mass-loading) parameter that measures the ratio of magnetic field to mass flux in the a frame that corotates with the footpoint of a field line labeled by the streamline $\psi$ averaged over X-wind streamlines, $\psi = 0$ to $\psi = 1$.  That is, ${\bf B} = \beta(\psi)\rho {\bf u}$ with $\bar\beta \equiv \int_0^1 \beta(\psi)\, d\psi$.  The
normalization for the streamfunction is the mass-loss rate in the wind, $\dot M_w$, which is a
fraction $f$ of the disk accretion rate $\dot M_D$.  The quantity $(GM_* \dot M_D^2 R_X^3)^{1/4}$
in equation (\ref{Xwind}) then provides the correct units when $GM_*$, $R_X$, and $\dot M_D$
are taken to be the fundamental dimensional quantities of the physical problem.

Setting $\Phi_t = 6\pi \bar B_h F_h R_*^2$ from equation (1), we obtain from equation (\ref{Xwind}):
\begin{equation}
F_h \bar B_h = \bar \beta f^{1/2} B_{\rm norm}\left( {R_X\over R_*}\right)^{3/4},
\label{Bhagain}
\end{equation}
where we have defined a fiducial field strength,
\begin{equation}
B_{\rm norm} \equiv \left( {GM_*\dot M_D^2\over R_*^5}\right)^{1/4}.
\label{Bnorm}
\end{equation}

We emphasize that the trapped-flux fractions, 1/3 for the funnel flow, 1/3 for the dead zone, and 1/3 for the X-wind, result from assuming that the thermal speed near the inner disk edge is small in comparison to its Kepler speed.  This assumption allows us to shrink the X-region to a mathematical point in the meridional plane, with the disk having infinitiesimal vertical thickness in the same approximation. How the vanishingly small diffusivities $\nu$ and $\eta$ then load field lines is not an addressable question in this limiting procedure, so we have to make a guess for this process.  For the somewhat ad-hoc loading law in the X-wind, $\beta = (3\bar(3\bar \beta/2)(1-\psi)^{-1/3}$, Table 3 of Cai et al. (2008) yields the identifications: $\bar J_w = 2.64, 4.36,$ and 6.20, respectively, for $\bar \beta = 1, 2,$ and 3.  With the further assumption that $f = 1/\bar J_w =1/3$ made for simplicity, we can then interpolate to obtain the identification that $\bar \beta = 1.21$.

\subsection{Models with Disk Locking}

We now impose the disk locking condition,
\begin{equation}
\Omega_* = \Omega_X = \left(GM_*\over R_X^3\right)^{1/2}.
\label{disklocking}
\end{equation}
With the stellar rotation rate given by the above, we obtain from equations (\ref{hotspotflux}) and (\ref{Xwind}) the desired general relationship:
\begin{equation}
F_h R_*^2 \bar B_h = \Phi_t = \bar \beta f^{1/2} (GM_*\dot M_D/\Omega_*)^{1/2}.
\label{JKG}
\end{equation}
Equation (\ref{JKG}), the exact analogue of a scaling relationship first derived by JG02, encapsulates the idea of flux trapping (see \S 2.5) because it relates the amount of measured flux in hot spots on the left-hand side to independently observable quantities
of the system on the right-hand side, without any assumptions about the multipolar character of the stellar magnetic field
ultimately responsible for the funnel flow.

\subsection{Dipole Models}

For the case of an undisturbed magnetic configuration for the star with a pure dipole of moment $\mu_*$, OS95 found the trapped stellar flux required for steady state to be
\begin{equation}
\Phit = {3\pi \mu_*\over R_X}.
\label{Ostriker}
\end{equation}
In the limit of vanishingly small resistivity, we can picture the trapped flux in equation (\ref{Ostriker}) to originate by the disk accretion flow sweeping all the exterior dipole flux from $\infty$ to $R_X$, $2\pi \mu_*/R_X$, into the X-region, plus another half that amount, $\pi \mu_*/R_X$, resulting from the interior dipole flux from $2R_X/3$ to $R_X$ being pushed into the X-region by the back reaction to the outward transport of angular momentum in the funnel flow.

Putting the expression (\ref{Ostriker}) into equation (\ref{Xwind}), we may solve for $R_X$ to obtain
\begin{equation}
R_X = \alpha_{\rm t} \left( {\mu_*^4\over GM_*\dot M_D^2}\right)^{1/7},
\label{Rx}
\end{equation}
where $\alpha_t$ is given by
\begin{equation}
\alpha_t = \left( {1\over 2\bar\beta f^{1/2}}\right)^{4/7},
\end{equation}
an expression first derived by OS95.  From the early computations of Najita \& Shu (1994), OS95 estimated $\bar \beta = 1$ would be needed to drive an X-wind with $f = 1/3$, giving $\alpha_t = 0.921$. (OS95 actually obtained 0.923 because eq.~\ref{Ostriker} represents a slight rounding of their numerical result for dipole funnel flows.) With the  more refined estimate $\bar \beta = 1.21$ for the case $f = 1/3$ from Cai et al. (2007) as discussed earlier, we get $\alpha_t = 0.825$ for the dipole model.  More general multipole models of funnel flows do not satisfy a simple relationship for the magnetopause standoff distance such as equation (\ref{Rx}); the dependence on the stellar dipole moment $\mu_*$ is replaced by a more complex dependence on the field geometry on the stellar surface at radius $R_*$.  This is encapsulated in the discussion of \S\S 3.3 and 4.2, where the surface field geometry helps to determine the quantities $F_h$ and $B_h$.

Although equation (\ref{Rx}) is reminiscent of similar expressions in Ghosh \& Lamb (1978, 1979a, b), we emphasize that the resemblance arises {\it mostly} as a result of dimensional analysis.  Given the three dimensional quantities $GM_*$, $\mu_*$ and $\dot M_D$, there is only one combination, $(\mu_*^4/GM_*\dot M_D^2)^{1/7}$, apart from a numerical coefficient $\alpha_t$ of order unity, that yields a quantity $R_X$ with the dimensions of length.  The physical reasoning underlying Ghosh \& Lamb's derivation for the disk truncation radius is quite different from those given above. Specifically, when steady state holds in a problem where the resistivity is small, the disk is nearly in mechanical equilibrium with stellar gravity balancing the centrifugal force of rotation.  Thus, there is no unbalanced ``ram pressure'' that the nearly {\it corotating} magnetic field at the disk's inner edge needs to offset.  What the stellar magnetosphere needs to do to truncate the disk is to transfer enough positive angular momentum outwards so as to offset the negative torques that are trying to drive the disk inwards in the equatorial plane.

The important point that disk truncation in CTTSs arises, not from ram pressure effects, but from considerations of angular momentum transport was first made by Cameron \& Campbell (1993).  Their analysis is closest in spirit to OS95's and ours.  The similarity is nearest in the prediction that disks are truncated at inner edges which lie close to the radius where the disk corotates with the central star, a phenomenon that has come to be called ``disk locking.''  Both analyses also ignore the presence of any intrinsic magnetization of the disk.  The main differences are that we suppose (1) that the disk resistivity is much smaller than the disk viscosity (by a factor proportional to the disk-aspect ratio; see Shu et al. 2007), and (2) that the viscous torque in the disk is negligible relative to the funnel-flow and X-wind torques, so that neither of the values of the coefficients of viscosity $\nu$ and resistivity $\eta$ enter the final formulae in the limit of a vanishing ratio for the sound speed to Keplerian velocity in the X-region, except for the dependence implicit in our assumption for the wind-loading fraction $f$.  In contrast, Cameron \& Campbell (1993) make detailed assumptions concerning $\nu$ and $\eta$, which they assume are equal in turbulent circumstances, and they ignore the X-wind torque, equating the funnel-flow torque to (the negative of) the viscous torque.  Moreover, instead of computing the funnel-flow torque in the funnel, they estimate it from the slip present between the mid-plane of the disk, assumed to rotate at Keplerian speeds, and its surface, assumed to corotate at the angular speed of the star, for regions that are magnetically connected to the star.  They parameterize this slip by an unknown factor $\gamma \ll 1$ that characterizes the resulting azimuthal magnetic field giving rise to a magnetic couple.\footnote{In the notation of this paper, equation (3) of Cameron \& Campbell (1993), which defines $\gamma$, reads
$${\left| {B_\varphi^+\over B_z}\right|} = {\gamma \over \eta}\varpi |\Omega_K-\Omega_*|z_0,$$
where $B_\varphi^+$ is the mean toroidal magnetic field at the surface of a disk of half-height $z_0$, and $\Omega_*$ and
$\Omega_K$ are, respectively, the rotation rates of the star and the disk (which is Keplerian) in the neighborhood of the truncation radius $\varpi$.  To avoid violent Parker instabilities, Cameron \& Parker require $|B_\varphi^+|$ to be not much larger than $B_z$, whereas their analysis suggests that $\Omega_K$ and $\Omega_*$ can differ by amounts that make $\varpi|\Omega_K-\Omega_*|$ considerably larger than the thermal speed $a$ in the midplane of the disk.  They consider two cases, where $\eta \sim az_0^2/\varpi$ and $\eta \sim az_0$ (cf. the unnumbered eq.~after eq.~3 and eq.~7 of their paper); in both cases, their equation (3) requires $\gamma \ll 1$ for disks that are spatially thin, i.e., for which $z_0/\varpi \ll 1$. Only a stretch of all parameter choices could make $\gamma \sim 1$.} 
If $\gamma \sim 1$, then their $R_m$ (= $R_X$ in our notation) can be larger than our equation (\ref{Rx}) by as much as a factor of 2 in typical circumstances (see \S 4.3).  Given the modeling uncertainties and the effects ignored in each exercise to obtain a tractable analysis, we do not consider this difference as numerically significant at the present state of observational tests.

In any case, the substitution of equation (\ref{Rx}) into equation (\ref{disklocking}) and $\mu_* \propto \bar B_h R_*^3$ results in a proportionality relationship specific to the dipole case,
\begin{equation}
\bar B_h R_*^3 \propto (GM_*)^{5/6}\dot M_D^{1/2} \Omega_*^{-7/6}.
\label{dipolerel}
\end{equation}
This is to be compared to the general expression for arbitrary stellar fields, equation (\ref{JKG}).

\subsection{Analysis by JG02}
JG02 have compared observations to the dipole model predictions by various groups, including that of OS95 embodied in equation (\ref{dipolerel}), as well as to the multipole prediction encapsulated in equation (\ref{JKG}).  To make progress, they assume that all T Tauri stars have the same value of $\bar B_h/\bar\beta f^{1/2}$, allowing both equations (\ref{JKG}) and (\ref{dipolerel}) to be treated as proportionalities.  In the restricted version of X-winds presented here, where $\bar \beta = 1.21$ and $f = 1/3$, this assumption amounts to postulating that $\bar B_h$ has the same value in all hot spots, a supposition that seems to hold approximately in many observed cases.  They find that the observations bear little correlation with all dipole predictions, including that of OS95.  The scatter is greatly reduced and a significant correlation is found, however, when the data are compared to the relationship of equation (\ref{JKG}), where the dipole approximation is dropped.  Thus, the generalized X-wind model for accretion flows is indeed compatible with current observations.

We note that the mean hot spot field strength, $\bar B_h$, is also in principle measurable by Zeeman effects.  Thus, if $\bar\beta f^{1/2}$ is a theoretically computable coefficient, then not only can the slope of a log-log plot of the variables on the left- and right-hand sides of equation (\ref{JKG}) be checked, but so can its intercept.  Indeed, $\bar B_h$ and $F_h$ have now been simultaneously measured in at least two T Tauri stars (V2129 Oph, Donati et al. 2007; BP Tau, Donati et al. 2008); we compare the data to our theory in \S 4.3 and find reasonable agreement.   


\section{Numerical Calculations}

We wish to pursue the consequences of the above theoretical discussion.  In particular, we wish to test whether flux-trapping at the X-point can lead to accretion funnel flows even when the unperturbed stellar field is not a pure dipole.  Our technique is to use direct numerical calculation, analogous to that carried out by OS95 for the dipolar case.  Specifically, we will model time-independent, axisymmetric accretion flows where the unperturbed field on the stellar surface is composed of higher-order multipoles.  We state at the outset that we adopt the simplification of OS95 that the sub-Alfvenic conditions of the funnel flow and the dead zones surrounding it can be approximated with vacuum fields plus current sheets, rather than treating the problem by the more accurate, but also considerably more complex, Grad-Shafranov equation (see, e.g., Cai et al. 2008).  Even with this simplification and the further assumption of axial symmetry, the removal of the dipole constraint means that our particular choices of the stellar field configuration will be neither unique nor universally applicable. We expect the actual field to differ from star to star, and to vary with time on any given star.  The goal of our study is more limited: to demonstrate explicitly how it is possible to satisfy the two primary observational constraints that the net surface field polarization in T Tauri stars is small, and that the accretion hot spots cover only a small part of the stellar surface.  


\subsection{Multipole Model}

The dipole model calculations for the inner (funnel-flow) solution have been discussed in detail by OS95.  Here we describe only our new multipole calculations for the funnel flow, while drawing attention to the salient points of departure from the dipole model.  The particular stellar multipole field we choose is discussed in the next section.

We assume an unperturbed stellar field that is some multipole aligned with the stellar rotation axis.  We also assume that the poloidal component of the final field (i.e., the perturbed field after accounting for interactions with the disk and flux-trapping) is in a vacuum-field potential configuration (i.e., is curl-free), {\it including} the field lines loaded with accreting matter.  This condition is approximately satisfied provided $(i)$ the Alv\'en Mach number of the flow, $\malven$, satisfies $\malven$ $\ll$ 1 everywhere, and $(ii)$ the net fractional specific angular momentum carried to the star in the funnel flow, $\jbstar$, satisfies $\mid \jbstar \mid$ $\ll$ 1.  The first condition means that the magnetic field energy dominates the matter kinetic energy even on the field lines participating in the funnel flow.  This is certainly true at both the X-point pinch and near the surface of the star, given the high field strengths in those regions. With the condition satisfied at both ends, we (like OS95) plausibly assume it is also true everywhere between.   The second condition means that any stellar spin-up or spin-down, due to the specific angular momentum carried on to the star by the accretion flow, is very slow - of order the accretion timescale - as is observed (see discussion of equation (2) in \S 2).  These considerations justify our (and OS95's) assumption of a curl-free poloidal field, $\del\times\vecBp$ = 0; i.e., the associated magnetic flux $\Phi$ satisfies the vacuum-field condition,
\begin{equation}
\del\cdot\left(\frac{\del\Phi}{{\varpi}^2}\right) = 0.
\label{vacuum}
\end{equation}

OS95 adopt the agnostic choice $\jbstar$ = 0 to keep a small star slowly rotating.  In this case, with $\malven$ $\ll$ 1, the field lines in the funnel flow must have a small but non-zero azimuthal component.  Combined with the poloidal field, this produces field lines that are trailing spirals, leading to an outward transfer of angular momentum, extracted from the inflowing material in the funnel flow by the resulting magnetic torques, and transferred back to the footpoints of the funnel flow field in the X-region of the disk.  Because the azimuthal component is small compared to the poloidal field in the funnel flow, the overall field geometry interior to the X-point is well represented by the poloidal field $\vecBp$ alone.  The latter is calculated by solving equation (\ref{vacuum}) subject to appropriate boundary conditions (described below).  The small azimuthal field $\vecBa$ can be found a posteriori, through the details of how the field lines are loaded with matter.        

Calculating the axisymmetric $\vecBp$ then reduces to a 2-D problem in the meridional plane.  Since we wish to solve only the inner problem (funnel-flow accretion), our calculations are confined to the region enclosed by the boundaries $z$=0 (axis along the equatorial plane, i.e., along the plane of the disk), $\varpi$=0 (rotation axis of the star and symmetry axis of the stellar multipole, perpendicular to the disk plane), $z$=$\zmax$ (upper boundary, parallel to the equatorial plane, of our modeled region) and $z$=$\zw$($\varpi$) (locus of the uppermost, i.e., innermost, wind streamline) (see Fig. 1).  The corresponding boundary conditions are: 

\begin{mathletters}
\begin{equation}
\Phi \rightarrow \Phim\,\,\,\,\,\, {\rm as} \,\,\,[{{\varpi}^2}+{z^2}] \rightarrow 0\;\;\;\;\;\;\;\;\;\;\;\;\;\;\;\;\;\;\;
\end{equation}
\begin{equation}
\Phi \rightarrow \Phif\,\,\,\,\,\, {\rm as} \,\,\,[{{(\varpi-\rx)}^2}+{z^2}] \rightarrow 0\;\;\;\;\;\;\;\,
\end{equation}
\begin{equation}
\Phi = \Phi_w = {\rm constant}\,\,\,\,\,\, {\rm on} \,\,\,z=\zw(\varpi)\;\;\;\;\;\;\;
\end{equation}
\begin{equation}
\Phi = \Phi_w(\varpi/{\varpi}_{\rm max})^2\,\,\,\,\,\, {\rm on} \,\,\,z=\zmax\;\;\;\;\;\;\;\;\;\;\;
\end{equation}
\begin{equation}
\Phi = \Phit\,\,\,\,\,\, {\rm on} \,\,\,z=0\,\,\, {\rm for} \,\,\,\rk\le\varpi\le\rx\;\;\;
\end{equation}
\begin{equation}
{{\partial}\Phi}/{{\partial}z} = 0\,\,\,\,\,\, {\rm on} \,\,\,z=0\,\,\, {\rm for} \,\,\,0\le\varpi<\rk
\end{equation}
\begin{equation}
\Phi = 0\,\,\,\,\,\,  {\rm on}  \,\,\,\varpi=0.\;\;\;\;\;\;\;\;\;\;\;\;\;\;\;\;\;\;\;\;\;\;\;\;\;\;\;\;\;\;\;\;\;
\end{equation}
\end{mathletters}
 
In the above, $\Phim(\varpi,z)$ is the unperturbed multipole flux; $\,\Phif \equiv \Phit$${\xangle}/{\pi}$ is the flux function for a fan of poloidal field over the angles $0\le\xangle\le\pi$ measured in the meridional plane with vertex at the X-point: $\xangle \equiv \arctan{[z/(\varpi - \rx)]}$.  The quantity $\Phit$ is the total amount of flux trapped between $\xangle$ = 0 and $\pi$; $\Phi_w = \Phi_t/3$ is the flux adjacent to the uppermost wind streamline.  The radius $\rk$ is the position of the ``kink point'' where stellar magnetic field lines make a transition from skimming horizontally over the equatorial plane for $\varpi > \rk$ to threading it vertically for $\varpi < \rk$, because the gas falls (in a trailing spiral) from the magnetic fan emanating from the X-point toward the star along the lowermost funnel-flow field lines that have been pulled into the inner edge of the accretion disk in back reaction to the torques exerted by the funnel flow. A more detailed explanation of the boundary conditions (16a)-(16b) is supplied in the Appendix, together with a prescription of how they are used in conjunction with the governing PDE (\ref{vacuum}) to find the magnetospheric configuration interior to the X-wind (see Fig. 1).

\subsection{Specification of Stellar Multipole Field}

Written in terms of a scalar potential $\Psi$, a poloidal field satisfies (Chandrasekhar 1961):
\begin{equation}
\vecB = \del \times \biggl[\del\biggl(\frac{\Psi}{r}\biggr) \times \vecr\biggr] .
\label{polfield}
\end{equation}
Imposing the curl-free condition $\del\times\vecBp$=0 reduces this to Laplace's equation for $\Psi/r$:
\begin{equation}
\delsq\biggl(\frac{\Psi}{r}\biggr) = 0.
\label{laplace}
\end{equation}
In a multipole expansion of spherical harmonics, this has the solution:
\begin{equation}
\Psi(r,\mu, \varphi) = \sum_{l=0}^{\infty}\sum_{m=0}^{\infty}A_{lm}\,r^{-l}\,B_{lm}\,r^{l+1}\,{Y_l}^m(\mu,\varphi),
\label{lapsolpsi}
\end{equation}
where $\mu$ $\equiv \cos\theta$.  We look for physical solutions wherein the strength of each multipole component of the unperturbed field decreases with increasing radius (distance from the star). We also wish to consider only azimuthally symmetric solutions independent of $\varphi$, i.e., with azimuthal quantum number $m$=0.  With no $\varphi$-dependence, the associated flux is 
\begin{equation}
\Phi_{\rm m}(r,\mu) = -(1-{\mu}^2)\frac{\partial\Psi(r,\mu)}{\partial\mu}.
\label{psiphi}
\end{equation}
Putting the above equations together, the unperturbed multipole flux can be expressed as 
\begin{equation}
\Phi_{\rm m}(r,\mu) = -(1-{\mu}^2)\sum_{l=0}^{\infty}A_l\,r^{-l}\,\frac{\partial{P_l(\mu)}}{\partial\mu},
\label{lapsolphi}
\end{equation}
where $P_l(\mu)$ are the usual Legendr\'{e} polynomials.  The function $\Phi_{\rm m}$ is known globally once it is fixed at any given radius $r$ = $R$ because the coefficients in equation (\ref{lapsolphi}) are
\begin{equation}
A_l = {\frac{(2l+1)}{2}}\,{R}^l\int_{-1}^{+1}\biggl[\int\frac{\Phi_{\rm m}(R,\mu)}{1-{\mu^2}}\,d\mu\biggr]\,P_l(\mu)\,d\mu .
\label{lapsolcoeff}
\end{equation}
Jardine, Cameron, \& Donati (2002) use a similar method to represent the surface magnetic fields of real stars derived from Zeeman tomography.

We elect to define the multipole flux $\Phi_{\rm m}$ at the stellar surface, $r$ = $\rstar \equiv 1$ for the present normalization.  To choose the surface flux, we note that the radial magnetic field on the surface is given by
\begin{equation}
B_r(\rstar,\mu) = \frac{1}{\rstar^2}\,\frac{\partial{\Phi_{\rm m}(\rstar,\mu)}}{\partial{\mu}} . 
\label{Brphi}
\end{equation}
We impose the conditions that the radial surface field is an odd function of $\mu$: $B_r(\rstar,\mu)$ = $-B_r(\rstar,-\mu)$, so that it vanishes at the equator: $B_r(\rstar,\mu)$ = 0 for $\mu$ = 0.  By equation (\ref{Brphi}), the same conditions must hold for $\Phi_{\rm m}'$ $\equiv$ $\partial\Phi_{\rm m}(\rstar,\mu)/\partial\mu$.  Up to an arbitrary multiplicative constant, a generic odd function that satisfies these conditions is
\begin{equation}
\Phi_{\rm m}'(1,\mu) = \mu^{2p+1}\,(1-\mu^2)^q,
\label{defphisurf}
\end{equation}
which reaches maximum/minimum (corresponding to maximum inward/outward directed $B_r$) at the latitudes $\mu_0$ = $\pm\sqrt{(2p+1)/(2p+2q+1)}$.  This form allows one to control both the compactness and the ingress/egress latitudes of the surface radial field, and hence of the final hot spots.  Multiple loops can be created by a superposition of such states with different exponents $p$ and $q$, and/or multiplying by a suitable oscillating even function in $\mu$.

For demonstration purposes, in this paper we use one such surface field:
\begin{equation}
\Phi_{\rm m}'(1,\mu) = {\rm cos}(8\theta)\,\mu\,(1-\mu^2).
\label{phisurfa}
\end{equation}
Expanding $\cos(8\theta)$ in terms of $\mu = \cos\theta$, we obtain
\begin{equation}
\Phi_{\rm m}'(1,\mu) = [1 - 32{\mu^2}\,(1 - \mu^2)\,(1 - 2{\mu^2})^2]\,\mu\,(1 - \mu^2)
\label{phisurfb}
\end{equation}

With $\Phi_{\rm m}'(1,\mu)$ specified, and the normalized radial scale $\rstar$ = 1, we have $\Phi_{\rm m}(1,\mu)$ = $\int\Phi_{\rm m}'(1,\mu)\, d{\mu}$. (The constant of integration can be set to zero without affecting the B-field). This expression is inserted into equation (22) to solve for the coefficients $A_l$, which are then employed in equation (21) to calculate $\Phi_{\rm m}(r,\mu)$ globally.  Note that equations (\ref{defphisurf})-(\ref{phisurfb}) express the non-dimensional flux on the stellar surface with an arbitrary scaling.  The dimensional value of the corresponding surface magnetic field depends on the stellar parameters, the accretion rate to be supported, and the details of how matter is loaded onto the field lines.  This physical scaling was discussed in \S2.2 (see equations (7) and (8)), and is made concrete for our specific models in \S4.2.      

Our choice of $\Phi_{\rm m}$, obtained from equation (\ref{phisurfa}), corresponds to 6 multipole components: $l$ = 1, 3, 5, 7, 9, 11, where $l$ = 1 is the dipole part.  The corresponding $A_l$, relative to the dipole component $A_1$, are given in Table 1.  Figure 2 shows the resulting $B_r$ and $B_{\theta}$ on the stellar surface and $B_{\theta}$ as a function of radius in the equatorial plane (note that $B_r$ = 0 in the equatorial plane). Figure 3 shows the unperturbed magnetic field lines over the entire computational domain.  The salient features of our chosen field are as follows.  

The $A_l$ are all roughly equal within a factor of a few, implying, via equation (\ref{lapsolphi}) with $r$ = $\rstar$ $\equiv$ 1, that the various components of the multipole flux are all roughly comparable (within an order of magnitude) on the stellar surface (the corresponding radial magnetic {\it field} on the surface, $B_r$($r$=1) is dominated by the higher orders, with a very weak dipole component).  The corresponding surface field geometry is highly loopy, as implied by the oscillatory shape of $B_r$ and $B_{\theta}$ in Figure 2 and explicitly illustrated in Figure 3.  {\it Thus the net polarization of the starlight derived from the surface field will be low.}  Moreover, since the radial field is confined to relatively narrow latitudinal bands, the self-consistent location of the X-point must be found so as to capture a significant fraction of the magnetic flux in these latitudinal bands; otherwise, there will not be enough trapped flux.    {\it Thus, the hot-spot covering fraction $F_h$ will also be low}.  Hence, highly loopy field configurations are  automatically consistent with the observed surface properties of CTTSs.  

Moving away from the star, the higher order multipoles rapidly weaken, with $B_l(r)$ $\rightarrow$ $r^{-(l+2)}$; the loopiness of the field decreases; and the purely dipole component finally dominates at large distances (at $\gtrsim$ 5$\rstar$ for our adopted field; see Fig. 3).  We can therefore expect the funnel flow solution near the X-point (i.e., the values of $\Phit/\Phimx$ and $\rk/\rx$) to resemble the OS95 dipole solution when $\rx$/$\rstar$ is sufficiently large.  However, we emphasize that, independent of the value of $R_X/R_*$, the important consideration is the trapped flux, and not the detailed multipole character of the stellar field.

It is noteworthy that the unperturbed field varies non-monotonically in both strength and direction with radius, unlike in the dipole case.  While each of the multipole components declines monotonically with radius, the vectorial directions of all the components are not the same.  Consequently, the multipole fields cancel out more in some regions than in others, yielding a summed field that oscillates in strength and direction with radius and is null at some radii, specifically, at 1.015$\rstar$ and 3.25$\rstar$ (Figs. 2, 3).  The summed field $\Phi_{\rm m}(r,\mu)$ finally decreases smoothly to zero at large distances, once the dipole dominates.  For the pure dipole case, OS95 normalized by $\rx$ to find a unique solution (in terms of $\Phit/\Phidx$ and $\rk/\rx$) {\it independent} of $\rx/\rstar$. This strategy is possible since the unperturbed field looks the same geometrically at all radii for a dipole (or for any other single higher order multipole), and the location of the stellar surface is immaterial for the scaled solution.  In our case, the location of the stellar surface is fixed by our choice of the normalization length for the specification of different multipoles.  As a consequence, the funnel flow solution ($\Phit/\Phimx$ and $\rk/\rx$) for a {\it specified} surface field configuration composed of a {\it mixture} of mutipoles cannot adapt unchanged to a stellar radius $R_*$ that is chosen arbitrarily relative to $R_X$.

However, once a funnel flow solution is found for a given $R_X$, the stellar surface can be redrawn at an arbitrary location to represent the solution for a {\it different} surface field than the one originally specified.  This is possible because, while the initial unperturbed field is fixed globally by its specified configuration on the stellar surface, our actual numerical calculation of the solution from this depends only on the location of $R_X$ relative to the origin and not with respect to the stellar surface (i.e., the stellar surface is not used as a boundary condition, the origin is).  Thus, while the derived solution corresponds to the originally specified surface field if the surface is drawn at $R_*$=1, it also corresponds to a different surface configuration for $R_*$$\ne$1 (with the relative surface field strength contributions of the $n$ multipoles in the new configuration being $\rstar^{-3}:\rstar^{-5}:...:\rstar^{n-3}$ times their originally specified values).  Obviously, the solution for a given $R_X$ will correspond to a different $R_X$/$\rstar$ for $\rstar$$\ne$1 compared to $\rstar$=1.  However, since the solution fixes the amount of trapped flux, the quantity ${F_h}{\bar B_h}{\rstar^2}$ -- the funnel flow magnetic flux on the stellar surface -- remains constant independent of the $\rstar$ location (see equation [1]).  In {\it dimensional} units, if $\rstar$ is held constant (and $M_*$ and $\dot M_D$ are fixed, so $B_{norm}$ remains unchanged), then the quantity $F_h$$\bar B_h$ is uniquely determined by $R_X$/$\rstar$ (see equation [7]), with each new redrawing of the stellar surface equivalent to a different $R_X$/$\rstar$.  What changes, given a solution at fixed $R_X$, is the spot covering fraction $F_h$ and corresponding $\bar B_h$ (with $F_h$ increasing as $R_X/\rstar$ decreases, since the funnel broadens with distance from the star, and $\bar B_h$ thus decreasing).  We shall use such redrawings to compare the dimensional results of our calculations to observations, in \S4.2 and 4.3. 

\subsection{Covering Fraction of Hot Spot}

For axisymmetric configurations, the covering fraction $F_h$ of the hot spot in the upper hemisphere is given by
\begin{equation}
F_h = \int_{\theta_1}^{\theta_2} \sin \theta \, d\theta = \mu_1-\mu_2,
\label{coveringfrac}
\end{equation}
where $\mu \equiv \cos\theta$ and $\theta_1$ and $\theta_2$ are, respectively, the upper and lower colatitudes of the hot-spot (annulus).  The numerical values of $\mu_1$ and $\mu_2$ are given by the flux tubes that link them to the upper and lower
surfaces of the trapped flux in the X-region, i.e.,
\begin{equation}
\Phi(r = R_*, \mu_1) = \Phi(r = R_X, \vartheta = \pi); \qquad \Phi(r = R_*, \mu_2) = \Phi(r = R_X, \vartheta = 2\pi/3).
\label{fluxlabels}
\end{equation}
When $R_X \gg R_*$, the hot spot is generally small, and we may approximate the flux difference on the stellar surface as arising from the unperturbed stellar field, $\Phi(R_*,\mu_2)-\Phi(R_*, \mu_1) \approx \Phi_m^\prime (R_*, \mu_h) (\mu_1-\mu_2)$, where $\mu_h = (\mu_1+\mu_2)/2$ gives the mean latitudinal location of the hot spot, and $\Phi_m^\prime (R_*,\mu)$ in unnormalized dimensionless form is the function (\ref{phisurfb}).  Moreover, the difference in $\Phi$ at $R_X$ between the angles $\vartheta = 2\pi/3$ and $\pi$ is simply 1/3 of the total trapped flux $\Phi_t$.  Thus, when the covering fraction is small, we have the approximation,
\begin{equation}
F_h = \mu_1-\mu_2 \approx {\Phi_t\over 3 \Phi_m^\prime (1, \mu_h)},
\label{approxFh}
\end{equation}
with it being unimportant in the ratio on the right-hand side whether we use scaled dimensional values for $\Phi_t$ and $\Phi_m^\prime (1,\mu_h)$ or unscaled nondimensional values.

\section{Numerical Results}
\subsection{Steady State Solutions}

We present dimensionless solutions for the X-point located at four representative radii: $\rx$ = 10$\rstar$, 7.5$\rstar$, 5$\rstar$ and 2$\rstar$.  In other words, we assume here that the dimensional parameters of the problem are such that the equilibrium value $R_X$ has the chosen ratio relative to $R_*$, before we consider in \S 4.2 the requisite value of the mean hot-spot field strength $\bar B_h$ needed to accomplish this feat.  The results are as follows.

\noindent
$\rx$ = 10$\rstar$ :  The final field configuration for this case is plotted in Figure 4.  The requirements are well met for a solution in which we may expect a sub-Alfv\'enic funnel flow to occur in the manner described by OS 95: the uppermost funnel flow field line just skims the equipotential curve near the X-point and never rises above this curve, ensuring steady accretion onto the star, and the lowest funnel flow field line subtends an angle of 120$^{\circ}$ to the disk as it exits at the kink point.  The accreting material lands on the star at a mean 
colatitude of $\theta = 42.0^\circ$.
The solution corresponds to $\Phit/\Phimx$ = 1.55 and $\rk/\rx$ = 0.70.  This is almost exactly the solution OS95 found for the dipole case: $\Phit/\Phidx$ = 1.5 and $\rk/\rx$ = 0.74.  As discussed above, this is to be expected, since the dipole component of the field vastly dominates at 10$\rstar$ even in our multipole model.  The situation is dramatically different close to the star, however.  Comparing OS95's dipole solution plotted in Figure 1 to our multipole one in Figure 4, we see that the flow in our model is squeezed into a much narrower funnel close to the star, as the higher order multipoles begin to dominate. Long et al. (2008) noticed a similar trend in their 3-D numerical simulations when the stellar magnetic configuration is complex.  In fact, the resolution of the inkjet printer used to make this figure is too low to show the true smallness of the hot spot; our hot spot covering fraction as computed by equation (\ref{approxFh}) is only 0.065\% (latitudinal width of only $\Delta\theta$ = 0.056$^{\circ}$)\footnote{In this and all other solutions below, the covering fraction computed from equation (29) is checked against that directly implied by the position of the first and last funnel flow field line on the stellar surface in the numerical models plotted in Figs. 4--7.  They agree in to within 0.01$^{\circ}$; the analytic values from equation (29) are quoted as the more accurate ones, given the finite spatial resolution of the numerical calculations.}, compared to the much larger $\sim$ 3\% for a pure dipole with $R_X$ = 10$R_*$ (see OS95).\footnote{Since our calculations assume axial symmetry, strictly speaking, we should speak in terms of ``hot rings'' rather than ``hot spots.''  However, the latter is the common usage among observers in the field, and we will follow this usage under the assumption that the qualitative concepts of the problem important to this paper will not change -- as is indeed demonstrated, for example, by the simulations of Long et al. (2007) -- if we were to perform a calculation that dropped the assumption of axial symmetry.}

\noindent
$\rx$ = 7.5$\rstar$  : The converged solution for this case is plotted in Fig. 5.  It corresponds to $\Phit/\Phimx$ = 1.71 and $\rk/\rx$ = 0.61.  The flow lands on the stellar surface at a mean colatitude of $\theta = 42.1^\circ$, with a covering fraction of 0.093\% ($\Delta\theta$ = 0.08$^{\circ}$), compared to $\sim$ 4\% for a pure dipole with $\rx$ = 7.5$\rstar$.  The solution parameters are close to the 10$\rstar$ case, but not identical, for the following reason.  While the dipole component utterly dominates at 7.5$\rstar$, the quadrupole starts to be non-negligible at smaller radii, with a polarity opposite that of the dipole.  This is evident in Fig. 5, where we see that $\rk$, the inner edge of the inward excursion of the disk, is not very far from the magnetic null point at 3.25$\rstar$, where the dipole field is completely cancelled by higher orders.  Thus the net flux swept into the X-region increases slower with decreasing radius than in the purely dipolar case, requiring a larger inward extension of the disk, i.e., smaller $\rk/\rx$, to trap sufficient flux than for a pure dipole. 

The above behavior raises the interesting question:  What happens as $\rx$ continues to approach the null point at 3.25$\rstar$ from radii exterior to it?   In the 7.5$\rstar$ case above, enough flux resides between the null point and $\rx$ to provide a good solution, despite the increasing importance of the quadrupolar component of opposite polarity to the dipole one.  At some stage, however, this will no longer be true, and flux from radii {\it interior} to the null point will start getting swept into the X-region exterior to it.  Since the field reverses at the null point, the net effect will be the reduction of trapped flux in the X-region.  A converged solution will then arise only when enough flux from radii interior to the null point has been swept in to offset this cancellation and once again provide sufficient trapped flux.  The field lines supporting the funnel flow must consist now entirely of ones that originate at low latitudes, below the second loop, instead of above it as in the 7.5$\rstar$ case (see Fig. 3).  The result is a latitudinal change of 30$^{\circ}$ in the hot-spot location (and a reversal in the polarity of the radial magnetic field in the hot spot compared to the previous location).  In other words, at some critical juncture, the hot-spot position must jump significantly for a small shift in $\rx$, or equivalently, for a small change in the accretion rate.  This is illustrated in the following solution.

\noindent
$\rx$ = 5$\rstar$  :  The final field configuration for this case is plotted in Fig. 6.  The solution parameters are $\Phit/\Phimx$ = -25.0 and $\rk/\rx$ = 0.247.  The flow lands at a mean colatitude of $\theta = 72.4^\circ$ on the stellar surface, with a covering fraction of 2.74\% ($\Delta\theta$ = 1.65$^{\circ}$), compared to $\sim$ 6\% for a pure dipole with $\rx$ = 5$\rstar$.  These numbers validate our thought experiment above.  The cancellation of flux in the X-region by field of opposite polarities requires far more flux from radii interior to the null point to be swept in to finally acquire enough trapped flux; consequently the inward excursion of the disk is much larger ($\rk/\rx$ much smaller), and the ratio of trapped to unperturbed flux ($\Phit/\Phimx$) much higher, than in the 7.5$\rstar$ and 10$\rstar$ cases above.  The negative sign of $\Phit/\Phimx$ denotes that the net trapped flux is oppositely directed to the original unperturbed flux at the X-point.  As a result, we see that the hot spot has now jumped to below the second loop, as predicted. 

The latter result resembles the conclusion reached observationally by Gregory et al. (2006) who deduced funnel streams flowing close to the equatorial plane when the central star has a complex magnetic topology.  The 3-D simulations of Long et al. (2007, 2008) also obtained matter to accrete onto hot spots close to the equatorial plane in such circumstances.

Rapid changes in hot spot location are indeed observed on CTTSs, a phenomenon impossible with a purely dipolar field (or any other single higher order multipole).  The discussion and calculation given above illustrate that such changes can only be explained by stellar fields that are complicated structures consisting of a mixture of multipoles.  Non-axial symmetry and time-dependence would add extra richness to this phenomenon.       

\noindent
$\rx$ = 2$\rstar$ :  The converged solution for this case is plotted in Figure 7.  The accreting gas now lands on the star at a mean colatitude of $\theta = 69.5^\circ$ with a covering fraction of 0.30\% ($\Delta\theta$ = 0.18$^{\circ}$) compared to the very large value of $\sim$ 21\% for a pure dipole with $R_X$ = 2$R_*$.  The solution parameters are now $\Phit/\Phimx$ = 40.0 and $\rk/\rx$ = 0.87; as in the 5$\rstar$ case, these are significantly different from the solutions at 7.5 and 10$\rstar$.  The reason again lies in the oscillatory behaviour of the magnetic field.  The X-point traps all the flux exterior to $\rx$, as well as some fraction of the flux interior to it.  Since the null point at 3.25$\rstar$ is exterior to $\rx$ = 2$\rstar$, many of the field lines swept into $\rx$, from radii both interior and exterior to it, cancel out.  Thus far more flux needs to be swept into the X-region (relative to the unperturbed flux residing there) from radii interior to it, to offset this cancellation and support a stable funnel flow, than at say $\rx$ = 10$\rstar$, where no such field turnover exists (either exterior to it or interior to it for radii $>$$\rk$).  This explains the much larger $\Phit/\Phimx$ in the 2$\rstar$ case.  At the same time, the field strength ($B_{\theta}$) in the equatorial plane rises much more rapidly, i.e., flux lines are much more tightly packed, with decreasing radius at 2$\rstar$ than at 10$\rstar$ (Figs. 2, 3).  Thus a relatively small excursion of the disk inward of $\rx$ = 2$\rstar$ is required to capture the requisite flux for a steady-state solution.  This accounts for the smaller $\rk/\rx$ at 2$\rstar$, in spite of the flux cancellation, than at 10$\rstar$.  The rapid increase in stellar field inward from $R_X
= 2R_*$ implies that it would be much harder to crush the stellar magnetosphere entirely (i.e., bring $R_X$ inside $R_*$ by increases of $\dot M_D$) than naive estimates with pure dipole formulae might suggest.

\subsection{Dimensional Field Strengths}

Typical numerical values for CTTSs are $M_* = 0.5$ $M_\odot$, $R_* = 2 R_\odot$, and $\dot M_D = 10^{-8} M_\odot$yr$^{-1}$.  Our range in simulated $\rx$, 2--10$\rstar$, then corresponds to disk-locked stellar rotation periods of 1.3 to 14 days, consistent with the observed span in CTTS periods.  These fiducial parameters also yield, via equation (\ref{Bnorm}), $B_{\rm norm} = 26.68$ G.  With $\bar\beta = 1.21$, $f = 1/3$, and $F_h$ = 3.0${\times}10^{-3}$, 2.74${\times}10^{-2}$, 9.3${\times}10^{-4}$ and 6.5${\times}10^{-4}$ computed for the cases $R_X/R_*$ = 2, 5, 7.5, and 10, equation (\ref{Bhagain}) then implies that the required mean strength for the hot spot field is $\bar B_h =$ 11, 2.3, 91, and 161 kG, respectively.  Because we have fixed the surface geometry, the absolute value of the field strength (without regard to sign), averaged over the entire stellar surface, is $\sim$60\% of the hot spot field in each case.   

The measured mean absolute field strengths on CTTSs are a few kG (e.g., Johns-Krull 2007).  Formally, this suggests that $R_X \approx 5 R_*$ is the expected standoff distance of the stellar magnetopause for the choice of field geometry and stellar and disk parameters ($M_* = 0.5 M_\odot$, $R_* = 2 R_\odot$, $\dot M_D = 1\times 10^{-8} M_\odot$) made here. The ratio $R_X/R_* \sim 5$ compares well as an average with the measured inner edges of CTTSs by infrared spectroscopy (e.g., Carr 2008).
However, the specifics of the field reversals in the $5 R_*$ model gives Figure 6 an especially large spatial extent from
$R_k$ to $R_X$, suggesting, perhaps, especially large X-rays flares when such regions undergo magnetic reconnection.  It would be interesting in this regard to map the magnetic field configurations on the stellar surface and their loops above the surface relative to the locations of the hot-spots of known YSO sources of large X-ray flares.  A pre-flare configuration that resembles Figure 6 and a post-flare configuration that resembles Figure 7 would be most suggestive of some of the mechanisms proposed in this paper for the reasons behind hot-spot jumps.

In contrast, the $\bar B_h$ we compute for $\rx$ = 7.5 and 10$\rstar$ are much too high, by almost 2 orders of magnitude, to represent realistic mean absolute field strengths on CTTSs.  Concurrently, the hot spots in these two cases, while smaller than for a dipole, as desired, are {\it too} small compared to observations.  Field configurations with large amounts of power in high-$l$ multipole components are unlikely to allow large magnetopause radii $R_X$ compared to $R_*$, for the following reason.  By construction, the X-region traps all flux from the outer edge of the disk to $\rk$.  Over this range of radii, the dipole component dominates in the equatorial plane for 10 and 7.5$\rstar$ (with the quadrupole component making a small contribution for 7.5$\rstar$).  On the stellar surface for these two cases, however, higher order multipoles (specifically, the $l$=5 and 9 components) characterize the hot spots.  In other words, the trapped flux that carries the funnel flow is predominantly dipolar in the X-region but overwhelmingly a superposition of multipoles of high order in $l$ at the hot spot.  For sufficient (dipole) flux to be trapped at $\rx$ = 7.5 and 10$\rstar$ to support a balanced X-wind/funnel flow, our (arbitrary) choice of the stellar field thus implies a total field in the hot spot, composed primarily of higher order multipoles, that is much larger than the dipole field alone and squeezes the funnel into an extremely small spot on the stellar surface.  The consequent small covering fractions $F_h$ and large required fields $\bar B_h$ in the hot spots then compare badly with the observations.  More suitable choices of the multipole distribution, combined with lower values of the mass-accretion rate $\dot M_D$ in the disk, can do better if relatively large values of $R_X/R_*$ are indicated (see \S4.3).  

Note however that the {\it product} $F_h\bar B_h$ in our numerical calculations is a reasonable number: 33, 63, 85 and 105 G respectively, in each of the cases $R_X/R_*$ = 2, 5, 7.5 and 10.  This suggests that in the context of our calculations, a simple remedy is to choose the actual stellar surface to be higher than the fiducial values drawn in Figures 4 and 5.  As discussed at the end of \S3.2, this is equivalent to {\it (1)} changing the adopted {\it surface} field configuration, and {\it (2)} concurrently changing the $\rx$/$\rstar$ that a particular solution (specified by $\Phit/\Phimx$ and $\rk/\rx$) corresponds to, while {\it (3)} leaving the solution parameters at the X-point ($\Phit/\Phimx$ and $\rk/\rx$) unchanged for the specified position of $\rx$ on the equatorial plane with respect to the origin.  In particular, a larger surface would both increase $F_h$ and decrease $\bar B_h$ (see end of \S3.2), rectifying the anomalous values derived for the two largest $R_X/R_*$ cases above.


To illustrate this, consider increasing the stellar radius by a factor of 10/7 in the $\rx$ = 10$\rstar$ case.  While the solution parameters at the X-point remain unchanged from the $\rx$ = 10$\rstar$ values ($\Phit$/$\Phimx$=1.55 and $R_k$/$\rx$=0.70), the flow now lands at a co-latitude $\approx$ 31.8$^{\circ}$, and {\it the X-point now effectively corresponds to $\rx$ = (10$\times$7/10)$\rstar$ = 7$\rstar$} (Fig. 4).  The new $B_r$ on the stellar surface and $B_{\theta}$ in the equatorial plane are shown in Fig. 2.  We see that the magnetic field on the new surface (in arbitrarily scaled, non-dimensional units) is lower by a factor of $\sim$20 compared to its original value; the contribution of the highest order multipoles has decreased significantly as well, with the quadrupole ($l$=3) and octupole ($l$=5) components now dominating.  The corresponding hot spot is noticeably larger (Fig. 4), with $F_h$ = 0.91\%.  Adopting the same $B_{norm}$ (i.e., unchanged dimensional stellar mass, radius and accretion rate), $\bar\beta$ and $f$ as above, equation (7) for $\rx$ = 7$\rstar$ with the new $F_h$ then implies $\bar B_h$ = 8.8 kG, much smaller than the 91 kG derived from our original surface field configuration for a similar $\rx/\rstar$ = 7.5.  The new $F_h$ and $\bar B_h$ are perfectly compatible with observed CTTs; indeed, they correspond quite closely to specific CTTs configurations, as discussed below.


\subsection{The Specific Cases of V2129 Oph and BP Tau}

Recently, Donati et al. (2007) have used spectropolarimetry techniques followed over the rotation cycle of the modestly accreting T Tauri star V2129 Oph to map the 3-D magnetic field on its surface.  They find that the funnel flow lands on the star in a complex,
{\it nonaxisymmetric}, pattern of hot spots that is distinctly non-dipolar in its geometry.  To minimize the derived stellar flux, they assume that the unseen hemisphere of the star is antisymmetric in its field distribution to the seen hemisphere.  They then reconstruct the multipolar field distribution by including spherical-harmonic $l$-values up to $l$ = 9, and find that the dominant contribution is from the the quadrupole and octupole components.  The observed dipole component corresponds to a field $B = 350$ G at the magnetic pole of the star, whose radius they estimate to be
$R_* = 2.4$ $R_\odot$.  Hence the star has a putative dipole moment $\mu_* = BR_*^3/2 = 8.13\times 10^{35}$ G cm$^3$.

If we adopt their data set, $M_* = 1.35 M_\odot$, $\dot M_D = 10^{-8} M_\odot$ yr$^{-1}$, and simply use the dipole formula for the magnetopause standoff distance (\ref{Rx}) with the OS95 estimate $\alpha_t = 0.923$, we get $R_X = 2.67 R_*$, which is appreciably smaller than the corotation radius of $R_{\rm co} = 6.67 R_*$ corresponding to a 6.53 day rotation period.  Part of the difficulty may lie in too large an assumed accretion rate for this fairly evolved pre-main-sequence star.  From a calibration of Paschen and Brackett line luminosities, Natta et al. (2006) obtain $\dot M_* = 5.5 \times 10^{-9} M_\odot$ yr$^{-1}$ for this object, consistent with the $4 \times 10^{-9} M_\odot$ yr$^{-1}$ Donati et al. also infer (but use as a lower limit) from the observed Ca II flux of V2129 Oph and the $\dot M_D$--Ca II flux relation derived for CTTS by Mohanty et al. (2005).  However, since we advocate a disk accretion rate that is $(1-f)^{-1} = 3/2$ times larger than the stellar accretion rate, the Natta et al. number translates to an $\dot M_D$ that does not differ much from the assumed $\dot M_D = 10^{-8} M_\odot$ yr$^{-1}$, and we shall continue to use this value as a standard of comparison.

With their input data, Donati et al. (2007) claim that the pure (aligned) dipole model of Cameron \& Campbell (1993) gives values of $R_m$ equal to $5.1 R_*$ (magnetic buoyancy model) to $8.1 R_*$ (turbulent model), in better formal agreement with the empirical value of $R_{\rm co}$ than the pure (aligned) dipole models of K\"onigl (1991) or OS95.  This claim is accurate; however, we note that the superior agreement results by choosing a value for the slip parameter $\gamma$ equal to unity, whereas footnote 1 explains that $\gamma$ is probably small compared to unity if we accept the estimates of Cameron \& Campbell at face value.  

The difference between pure dipole and more complex multipole models provides a better resolution.  Suppose that we can apply the axisymmetric equations of \S 2.2 to V2129 Oph.  For $M_* = 1.35 M_\odot$, $\dot M_D = 1\times 10^{-8} M_\odot$ yr$^{-1}$, and $R_* = 2.4 R_\odot$, equation (\ref{Bnorm}) yields $B_{\rm norm}$ = 27.2 G.  With $\bar\beta =1.21$, $f = 1/3$, and $R_X/R_* = 6.67$, equation (\ref{Bhagain}) yields $F_h\bar B_h = 79$ G, in reasonable agreement with the observed value $F_h\bar B_h \approx 100$ G (where $F_h$ $\approx$ 5\% and $\bar B_h$ $\approx$ 2 kG is the mean strength in the hot spots of the total field, not just the dipole component).

Donati et al. (2008) have carried out similar observations of the CTTS BP Tau, with parameters $M_*$ = 0.8$M_\odot$, $\rstar$ = 1.95$R_\odot$, $\dot M_D$ $\approx$ 3 $\times$ 10$^{-8}$$M_\odot$ yr$^{-1}$, and rotation period $P$ = 7.6 days corresponding to $\rx/\rstar$ = 7.5.  They find a strong, mainly axisymmteric poloidal field, with roughly equal mean field strengths of 1.2 kG and 1.6 kG respectively in a dipole and an octupole component (both slightly tilted with respect to the rotation axis).  From the stellar mass, radius and accretion rate, equation (8) gives $B_{norm}$ = 53.7 G.  With adopted $\bar\beta$ = 1.21 and $f$ = 1/3 as before, equation (7) then predicts $F_h$$\bar B_h$ = 170 G, in excellent agreement with the data: Donati et al. measure a mean hot spot field strength $\bar B_h$ $\approx$ 9 kG (including both dipole and octupole components) and covering fraction $F_h$ $\approx$ 2\%, leading to an observed $F_h$$\bar B_h$ $\approx$ 180 G.  


It is also noteworthy that the V2129 Oph and BP Tau observations resemble our scaled calculation presented at the end of the last section (\S4.2).  In both the model and the data, the surface fields are dominated by multipole components up to an octupole, the $\rx$ are similar (6.7--7.5$\rstar$ in the data, intermediate 7$\rstar$ in the model), and the flow lands at high latitudes; under the circumstances, it is heartening that (with adopted $R_*$, $M_*$ and $\dot M_D$ comparable to the data) the predicted $F_h$ ($\sim$1\%) and $\bar B_h$ (8.8 kG) are also quite similar to the observed values (2--5\% and 2--9 kG respectively).  While we have made no attempt to exactly reproduce the data (and indeed cannot, given our assumptions of axisymmtery and aligned field), it is thus evident that both our analytic mutipole equations and  corresponding numerical models can reproduce the broad qualitative and quantitative features of the data.

Thus, the time has probably come to abandon formulae for disk truncation radii based on simple dipole models for the unperturbed stellar field in YSOs.  We take encouragement that progress in both the observational and theoretical developments allow researchers to argue meaningfully about missing factors of 2 or so in as complex a subject as the magnetospheric-disk interactions of CTTSs.

On the other hand, the 3-D simulations of Long et al. (2007, 2008) always disrupt the disks at a point where ``ram pressure balance'' holds approximately (see also the calculations of R03, R04, Long et al. 2005).  However, as discussed in \S 1, none of these simulations produce an X-configuration of trapped flux.  Trapped flux creates fields near the inner edge of the disk that depart appreciably from an unperturbed stellar dipole.  Thus, until simulations of the type performed by R08 are carried out with complex (non-dipolar) fields on the stellar surface, we regard factors of 2 uncertainties in the location of the stellar magnetopause as an open question.

As a final comment, we should note that although the global field geometries reconstructed as Figs. 15-16 of Donati et al. (2007) and Fig. 15 of Donati et al. (2008) are visually compelling, the extrapolation of exterior magnetic fields from measured ones on a stellar surface is, in general, a mathematically ill-posed procedure and does not produce unique configurations.  As an example, the boundary condition (16a) of this paper does not suffice, {\it by itself}, to fix the solution of the elliptic PDE (\ref{vacuum}) for $r > R_*$.  Also needed are the other conditions (16b)-(16g).  Other models for the disk-magnetosphere interaction will involve different outer boundary conditions and/or singular surfaces, lines, and points. The combination of empirical measurements of the 3-D fields on the stellar surface, accurate theoretical modeling and better determinations of accretion rates is needed to produce further progress in the field.

\section{Conclusions} 

Since its inception more than a decade ago, generalized X-wind theory has made many successful predictions and survived several  key confrontations with observations (e.g., Johns \& Basri 1995; JG02, Shang et al. 2004;  McKeegan 2006; Pyo et al. 2006; Zolensky et al. 2006; Carr 2008; Edwards 2008). Claims of large rotation rates seen in YSO jets have been used to argue against X-winds (e.g., Bacciotti et al. 2002, Coffey et al. 2004).  These claims turn out to have difficulties, e.g., indications in some cases that the inferred rotation in the jet is counter to that of the disk, or the absence of any obvious rotation in jets best oriented (i.e., in the plane of the sky) to show such rotation (e.g., Cabrit et al. 2006, Pety et al. 2006, Coffey et al. 2007, Lee et al. 2006, 2007).  Present upper limits on the rotation in the latter cases show that if any disk wind is present in the observations, they must have small launch radii that make them look very similar to X-winds (see Cai et al. 2008 for a fuller discussion).  In a similar manner, the finding that the magnetic fields on the surfaces of CTTSs are locally strong as seen in hot spots, but globally weak when averaged for the photospheric polarization signal (Valenti \& Johns-Krull 2004), turns out not to be an argument against disk-locking via funnel flows (Matt \& Pudritz 2005), but evidence for the phenomenon of flux trapping predicted by X-wind theory (OS95) and quantified in this paper (see also JG02).  

In hindsight, we should perhaps not be surprised that fully convective CTTSs have surface distributions of magnetic field that are non-dipolar.  Such configurations complicate the theoretical modeling of funnel flows, but they break the degeneracy in the very different explanations offered by Ghosh \& Lamb (1978, 1979a, b; see also K\"onigl 1991) and by Cameron \& Campbell (1993), Shu et al. (1994a),
OS95, and this paper for the standoff distance $R_X$ of the stellar magnetopause where the disk is truncated at its inner edge and to whose Keplerian rotation the star is locked in steady state.  In particular, this paper shows that the observed field strengths in funnel-flow hot spots and their fractional covering fractions $F_h$ have a rational explanation in the context of generalized X-wind theory, while they are inexplicable in other semi-analytic formulations of the steady-state problem. The strong concentrations of magnetic flux available to realistic configurations of surface fields on young stars may mean that it is harder to crush their magnetopsheres than has been estimated by naive dipole estimates.

While the results from numerical simulations (e.g., Goodson et al. 1999, K\"uker et al. 2003, Long et al. 2005) contain many of the elements seen in the semi-analytic theory, complete agreement has not yet been achieved, perhaps because the numerical simulations invariably contain, so far, too much resistive diffusion of the field, relative to angular momentum transport by turbulent viscosity, to give a good semblance of the phenomenon of trapped flux (for progress in this regard, see Romonova et al. 2008, p. 281).  Nevertheless, the trends are promising, and at some stage, the awesome computing power of modern machines will be able to reliably extend the solutions to the non-axisymmetric and time-dependent regimes occupied by actual systems that are inaccessible to semi-analytic techniques.

\acknowledgements
We thank Suzan Edwards for informative conversations, Hsien Shang for providing Figure 1 and calculations of the wind interface, and the anonymous referee for detailed comments that greatly improved the paper.  SM acknowledges support from a {\it Spitzer Fellowship}; FS, from a NSC grant awarded to TIARA in Taiwan, where this work was completed but not written up until both authors were back in the USA.

\appendix
\section{Solution of Governing PDE with Associated Boundary Conditions}

We wish to solve the elliptic PDE (\ref{vacuum}) subject to the boundary conditions (16a)-(16g).
Condition (16a) requires that the field approach a sum of aligned multipoles near the origin.  The goal of our analysis is to mimic the observed characteristics of T Tauri stellar surface fields (\S3.2). The required sum is thus found by specifying the unperturbed multipole field at a finite radius $\rstar$ from the origin.  A non-restrictive multipole sum has the advantage that it defines an analytically differentiable field over all space excluding the origin (see \S3.2).  This allows us to perform the numerical computation without the added complexity of introducing the curved stellar surface as a boundary.  The actual location $R_*$ of the stellar surface is important, however, for astrophysical interpretations (see \S 3.2).

Condition (16b) demands that the field take the shape of a complete fan near the X-point, where a specified amount of flux $\Phit$ is trapped.  OS95 show that a fan is the simplest curl-free solution for trapping flux at a single point. 

Condition (16c) says that the fields under consideration are confined by the X-wind, which cannot be approximated accurately as a vacuum field. Thus, the uppermost wind streamline corresponds to the flux tube just beyond $\Phi$=$\Phi_w$.  All the field lines participating in the wind (as well as in the funnel flow) have their disk footpoints at the X-point, where the field assumes the shape of a fan.  Moreover, for a cold flow, the wind streamlines must exit the X-point at a shallower angle to the disk than the critical equipotential, in order for outflowing material to be successfully loaded onto them.  The critical equipotential on the wind side of the flow is inclined at $\vartheta =$ 60\deg to the disk at the X-point.  Thus, we take the {\it uppermost} wind streamline $\zw$($\varpi$) to attach to the X-point at an angle of exactly 60\deg to the disk.  These considerations enable us to identify the constant flux along $\zw$($\varpi$) as $\Phi_w$ = $\Phif$ = $\Phit\xanglew/\pi$, where $\xanglew \equiv \pi/3$.  

H. Shang (2005, private communication) kindly provided us the locus of the wind interface $\zw$($\varpi$) that she obtained from a detailed numerical calculation of the force balance across the dead-zone/X-wind interface, as elucidated by Shu et al. (1995).  This locus has a slope $=$ tan[$\pi$/3] (= $\tan \xanglew$) near the X-point, and approaches a vertical orientation more slowly than OS95's more arbitrary parametrization.  We fit a quadratic polynomial to the numerically obtained locus by $\zw(\varpi)$ = $a\,({\varpi/\rx})^2 + b\,({\varpi/\rx}) + c$ where $a$ $=$ 0.197 and $b$ $=$ 1.333, with $c$ = $-(a+b)$ guaranteeing that the streamline intercepts the disk, $z$=0, at the X-point $\varpi$ = $\rx$.   


Condition (16d) arises from the following considerations.  Since OS95 adopted an analytic expression for $\zw(\varpi)$ that can be extended to infinity, they were free to choose an upper boundary to their computational region at $\zmax=\infty$, and assume there the simple boundary condition $\partial \Phi/\partial z = 0$.  In our case, the locus $\zw(\varpi)$ from Shang is numerically calculated and thus forced to be spatially finite. Within our computational domain -- the region interior to the wind interface -- we therefore replace the OS95 condition of a uniform distribution of longitudinal field at $\zmax=\infty$ by one at a large but finite $\zmax = 15$. The flux function that gives a uniform value for $B_z$ is $\Phi = \Phi_w (\varpi/\varpi_{\rm max})^2$, where we have imposed the boundary values that $\Phi = 0$ on the axis $\varpi = 0$ and $\Phi = \Phi_w = \Phi_t/3$ at $\varpi = \varpi_{\rm max}$ when we reach the wind interface at the height $z=\zmax$.

If all the disk material came to an abrupt end at the X-point, then field lines fanning out from the X-point down to $\vartheta$ = $\pi$ (i.e., down to the midplane interior to the X-point) would reconnect with their oppositely directed counterparts across the midplane.  To prevent this, we (like OS95) relax the constraints to allow a current sheet (later called ``the reconnection ring'') to extend slightly inward of $\rx$, to the radius $\rk$. The latter, the inner terminal point of a current sheet (which extends in its exterior into the disk for $\varpi > R_X$), is called the ``kink-point'' radius and is the vertex of a magnetic ``Y'' if we consider the corresponding field geometry below the mid-plane (see below).  The last horizontal fan flux-tube from the X-region (both above and below the midplane), with value $\Phi$ = $\Phit$ and carrying the terminal part of the funnel flow, is then able to skim the midplane between $\rx$ and $\rk$ before bifurcating into two lines that form the innermost boundaries of the out-of-plane funnel flow. This bifurcation encapsulates the statement contained in condition (16e).  

Inward of $\rk$, the inspiraling material is forced out of the disk plane. The magnetic lines of force encountered in the midplane at these radii are closed multipole fields with footpoints in the star but not the disk.  For decreasing radii, they become increasingly unperturbed as the extension of the vacuum fields above the star.  Reflection symmetry across the midplane then gives rise to condition (16f).  Note that conditions (16e) and (16f) imply that the field at the midplane is completely vertical for $\,\varpi<\rk\,$, and completely horizontal for $\,\varpi>\rk$.  The field at $\rk$ must then have a kink, as illustrated in Figure 1.  We see that $\Phi$ = $\Phit$ thus represents the last field line emanating from the star that is pinched into the X-region, with a long slender beak extending from $\rk$ to $\rx$.  OS95 demonstrate that near $\rk$, the ideal geometry of the magnetic field lines in the meridional plane is an even three-way split of the full angle 2$\pi$, among the three magnetic sectors - field lines leading towards the X-point, emanating from the X-point, and unconnected to the disk - that have equal field strengths across their interfaces (Fig. 1).  Thus, the last flux line connected to the X-point, $\Phit$, subtends an angle of 120\deg as it leaves the midplane at $\rk$ (creating the kink).  In our (and OS95's) formulation, the kink radius $\rk$ represents an eigenvalue of the problem, whose value must be determined as part of the solution for $\vecBp$.  In other words, $\rk$ must be specified such that the derived $\vecBp$ conforms to the physically valid field geometry near $\rk$ described above.   

Finally, boundary condition (16g) sets the zero of the flux-function along the z-axis for axial symmetry. Thus, the field line exiting the pole of the star ($\Phi$ = 0) attaches to its open counterpart in the wind region only at $z=\infty$. 

Conditions (16c)--(16g) specify $\Phi$ or its normal derivative on a closed curve. For a 2-D elliptic PDE in $\Phi$ such as equation (\ref{vacuum}), these alone usually suffice to completely determine $\Phi$ throughout the bounded region.  Yet we (and OS95) are able to impose, apart from the midplane Y (which has an associated eigenvalue), two further conditions, (16a) and (16b), at the origin and X-point respectively, without over-determining the problem.  This is because the required solution for $\Phi$ is not differentiable at the origin (singular dipole for OS95, singular multipoles for us) or at the X-point (magnetic fan). The multipole and fan behaviours must be imposed explicitly in order to obtain the desired solution that implicitly has current sheets and divergent current loops at the boundaries and the origin of the problem.  To deal with these singularities at the origin, X-point, and midplane Y, we follow OS95 in solving the PDE as follows.  

We define
\begin{equation}
\Phi = \Phim + \Phif + \Phir ,
\label{defphi}
\end{equation}
where $\Phim$ and $\Phif$ are the already defined multipole and fan flux functions; $\Phir$ is the remainder field that must be computed to completely determine the final solution $\Phi$.  We also define the differential (Stokes) operator, 
\begin{equation}
\dop \equiv \varpi\frac{\partial}{{\partial}\varpi}\left(\frac{1}{\varpi}\frac{\partial}{{\partial}\varpi}\right) + \frac{{\partial}^2}{{{\partial}z}^2},
\label{defdop}
\end{equation}
so that the PDE of equation (\ref{vacuum}) is equivalent to $\dop$($\Phi$) = 0.  Since the multipole field $\Phim$ itself obeys $\dop$($\Phim$) = 0, the remainder field must, by equation (\ref{defphi}), satisfy
\begin{equation}
{\dop}(\Phir) = -{\dop}(\Phif).
\label{solPDE}
\end{equation}
Combining the earlier boundary conditions on $\Phi$ (eqs. 16a--g) with equation (\ref{defphi}), we derive the corresponding boundary conditions on $\Phir$ to be 

\begin{mathletters}
\begin{equation}
\Phir = -\Phim\,\,\,\,\,\, {\rm on} \,\,\,\zw(\varpi)\;\;\;\;\;\;\;\;\;\;\;\;\;\;\;\;\;\;\;\;\;\;\;\;\;\;\;\;\;\;\;\;\;\;\;\;\;\;\;\;\;\;
\end{equation}
\begin{equation}
\Phir\,=\,\Phif(\varpi/{\varpi}_{\rm max})^2\, - \,\Phif\, - \,\Phim\,\,\,\,\,\, {\rm on} \,\,\,z=\zmax\;\;\;\;\;
\end{equation}
\begin{equation}
\Phir = -\Phim\,\,\,\,\,\, {\rm on} \,\,\,z=0\,\,\, {\rm for} \,\,\,\rk\le\varpi\le\rx\;\;\;\;\;\;\;\;\;\;\;\;\;\;
\end{equation}
\begin{equation}
\partial\Phir/\partial{z} = -\partial\Phif/\partial{z}\,\,\,\,\,\, {\rm on} \,\,\,z=0^+\,\,\, {\rm for} \,\,\,0\le\varpi<\rk
\end{equation}
\begin{equation}
\Phir = -\Phif\,\,\,\,\,\, {\rm on} \,\,\,\varpi=0\;\;\;\;\;\;\;\;\;\;\;\;\;\;\;\;\;\;\;\;\;\;\;\;\;\;\;\;\;\;\;\;\;\;\;\;\;\;\;\;\;\;\;
\end{equation}
\end{mathletters}


We are interested in the geometry of the total field. Because the governing PDE is linear and equidimensional, the geometry is unaffected by spatial or field-strength scaling.  All spatial parameters are thus normalized by $\rstar$, and field-strengths by $\Phimx$, the unperturbed multipole flux at the X-point in the absence of interactions with the disk.  That is, the problem is scaled so that $\rstar$ $\equiv$ 1 and $\Phimx$ $\equiv$ 1.  With the above boundary conditions, solving equation (\ref{solPDE}) for the scaled $\Phir$ then requires only a specification of two quantities: (i) $\Phit/\Phimx$, the ratio of flux trapped at the X-point to the unperturbed multipole flux there in the absence of a disk; and (ii) $\rk/\rx$, the ratio of the kink point radius to the X-point radius. The value of $\Phit/\Phimx$ is determined as the minimum required to have no ``frustrated'' field lines for steady funnel flow (i.e., no field line leaves the X-point on the ``downhill'' side of the critical equipotential, yet subsequently crosses it again so that a cold funnel flow would encounter an insurmountable potential barrier; see OS95), while the eigenvalue $\rk/\rx$ is found by trial and error, given $\,\Phit/\Phimx\,$, by requiring the solution to satisfy the ideal Y-field geometry near $\rk$.  The derived remainder field $\Phir$ is then superimposed on the (scaled) $\Phim$ and $\Phif$ to obtain the total scaled solution $\Phi$ from equation (\ref{defphi}).  

To solve equation (\ref{solPDE}) and its associated boundary conditions on a grid, we use the transformation ($\varpi$,$z$) $\rightarrow$ ($\varpi$,$\zeta$), where $\zeta$ $\equiv$ 2$\lambda$[$\sqrt{(z/\lambda)-1} - 1$].  Here $\lambda$ $\equiv$ ${(2a + b)^2}/4a$, where $a$ and $b$ are the coefficients of our quadratic form for the wind interface defined earlier.  This transformation maps the original curved wind interface $\zw(\varpi)$ = $a\,{\varpi}^2 + b\,{\varpi} + (a-b)$ to the inclined straight line $\zeta(\varpi)$ = 2$\sqrt{a\lambda}$[$\varpi - 1$], yielding a trapezoidal grid.  The PDE is then solved by simultaneous overrelaxation through finite-differencing.

\clearpage

\plotone{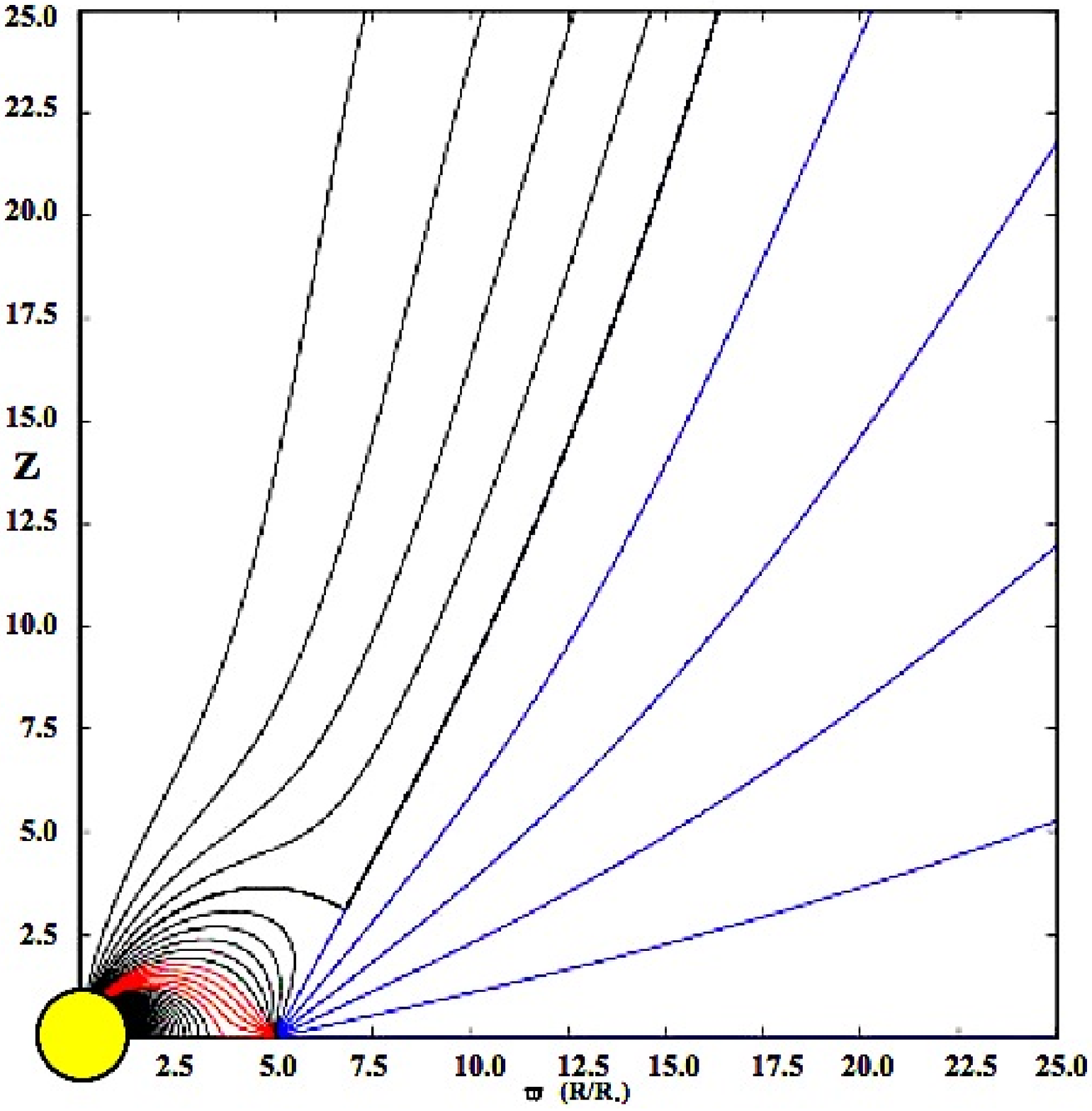}
\figcaption{Original X-wind accretion/outflow solution, using a stellar dipole field, that combines the funnel flow of OS95 with the X-wind solution of Shu et al. (1995) for the case $\rx$ = 5$\rstar$ (modified version of figure kindly supplied by H. Shang).  {\it Solid curves} are magnetic field lines: accretion funnel flow lines in {\it red}, open field lines from the X-point in {\it blue}, all others in {\it black}.  Key features generic for good solutions: flux-trapping at the X-point $\rx$, extension of an infalling disk slightly inward of $\rx$ until the kink-point $\rk$ which makes an angle of 120$^\circ$, and the uppermost funnel flow field line always remaining below the equipotential curve. But notice the high net field polarization on the stellar surface, and the relatively large hot spot covering fraction ($\sim$6\% if $\rx/\rstar = 5$) implied by the dipole field. See \S\S 2 and 3.1}  
\plotone{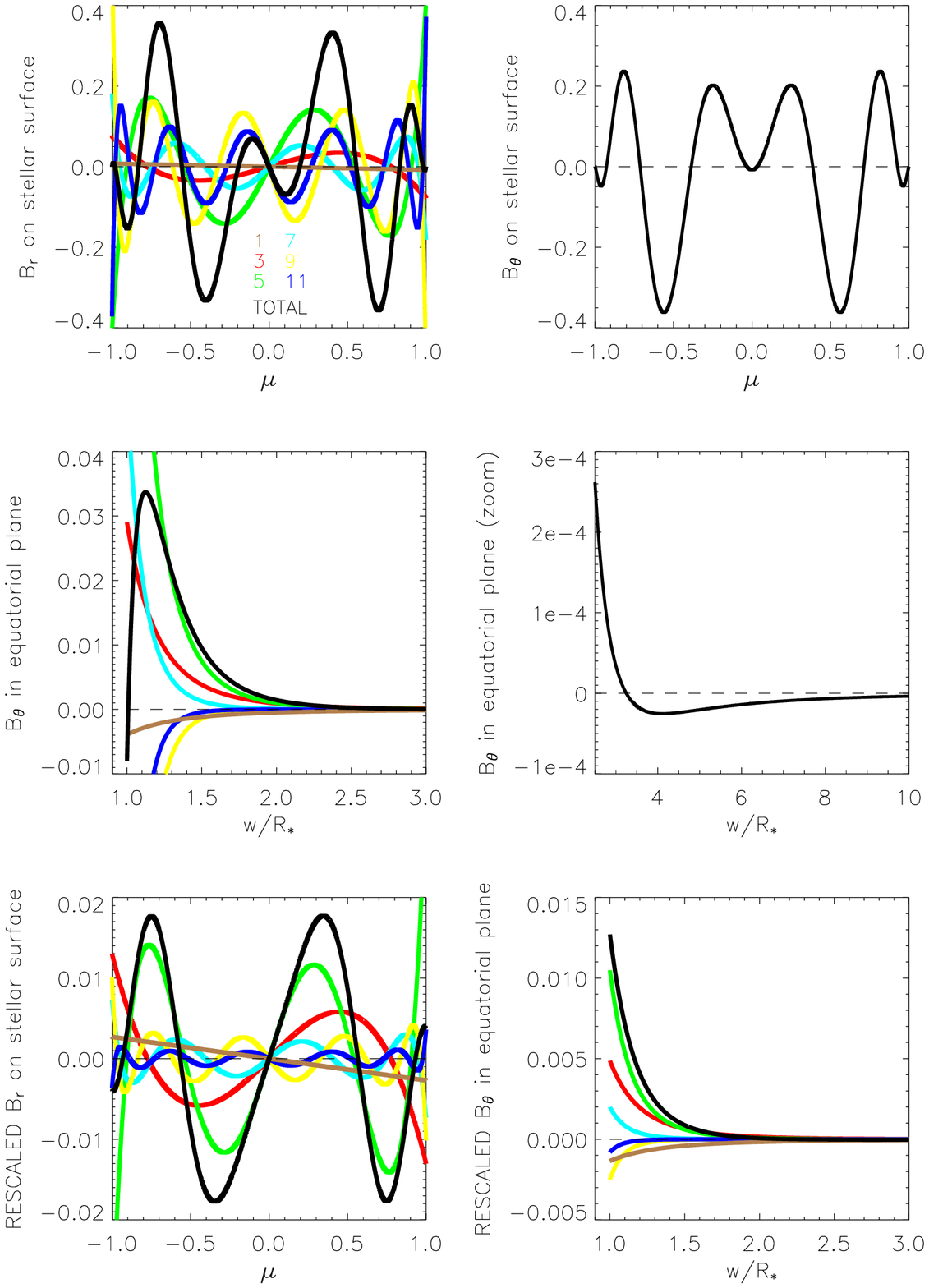}
\figcaption{The unperturbed stellar magnetic field ($B_r$ and $B_{\theta}$) corresponding to the $\Phim$ of equation (25), in non-dimensional, arbitrarily scaled units (but same relative scaling in all panels).  {\it Top left}: $B_r$ on stellar surface, as a function of $\mu$.  The various multipole components $l$ = 1--11 are shown as {\it colored} lines (see key in plot); the corresponding total $B_r$ is in {\it black}.   {\it Top right}: Total $B_{\theta}$ on stellar surface as a function of $\mu$ (individual multipole components not shown).  Both $B_r$ and $B_{\theta}$ oscillate rapidly, signifying a loopy field structure on the surface.  {\it Middle left}: $B_{\theta}$ as function of radius (in units of stellar radius) in the equatorial plane (with $\rstar$ $\equiv$ 1; $B_r$ = 0 in this plane).  Individual multipole components in {\it color} (same key as in top left panel), total in {\it black}.  {\it Middle right}: Zoom-in of total $B_{\theta}$ in the equatorial plane at large radii.  Note that $B_{\theta}$ changes direction at 1.01$\rstar$ and 3.25$\rstar$ (passing from negative to positive values), finally decreasing smoothly to zero at large radii as the dipole field dominates.  {\it Bottom left}: $B_r$ as a function of $\mu$ on rescaled stellar surface, $R_*$ = 10/7 (as opposed to $R_*$ = 1 in the previous panels).  {\it Bottom right}: $B_{\theta}$ as a function of radius (in units of stellar radius) in the equatorial plane, for the same rescaled stellar surface.  Multipole components in both panels in {\it color} (same key as in top left panel), total field in black.  From comparison to top left and middle left panels, note that (1) $B_r$ and $B_{\theta}$ are $\sim$20 times weaker for the larger rescaled surface compared to the smaller original one, and (2) the quadrupole and octupole ($l$ = 3 and 5) components largely dominate for the rescaled surface. See \S\S 3.2 and 4 (particularly \S4.2 for the rescaled surface). }
\plotone{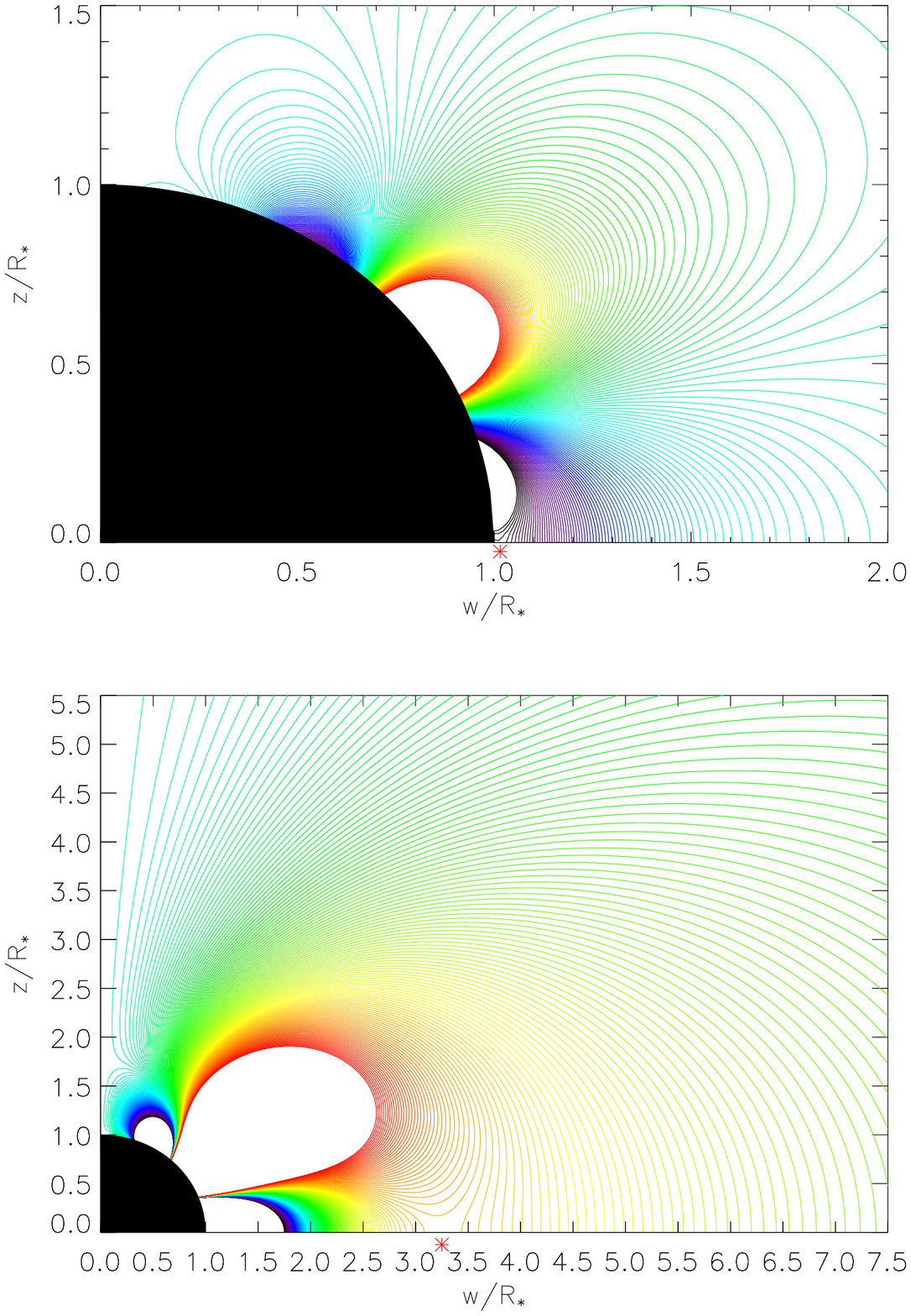}
\figcaption{Magnetic lines of force in the meridional plane, corresponding to the $\Phim$ of equation (25).  Note that this $\Phim$ is non-dimensional and arbitrarily scaled; absolute scaling and dimensional field strengths are discussed in \S4.2.  {\it Top panel}: Zoom-in of field close to the star.  Contours trace lines of constant flux, from $\Phim$  = -0.035 to +0.035, at intervals of 0.0005.  The field is highly loopy near the stellar surface.  {\it Bottom panel}: Same as above, except over a larger spatial range.  Contours are from -0.002 to +0.002, at intervals of 0.00002 (1/15 of the range and 1/25 of the spacing in the top panel).  Far from the star, the field increasingly resembles a simple dipole.  {\it Colors} denote contour levels: {\it blue} $\rightarrow$ {\it red} $\Rightarrow$ -ve $\rightarrow$ +ve.  The field direction thus corresponds to progression in colours with $r$ or $\theta$ (for $B_{\theta}$ and $B_r$ respectively).  {\it Red asterisks} mark the two null points in the field in the equatorial plane, where $B_{\theta}$ changes direction. See \S\S 3.2 and 4. }
\plotone{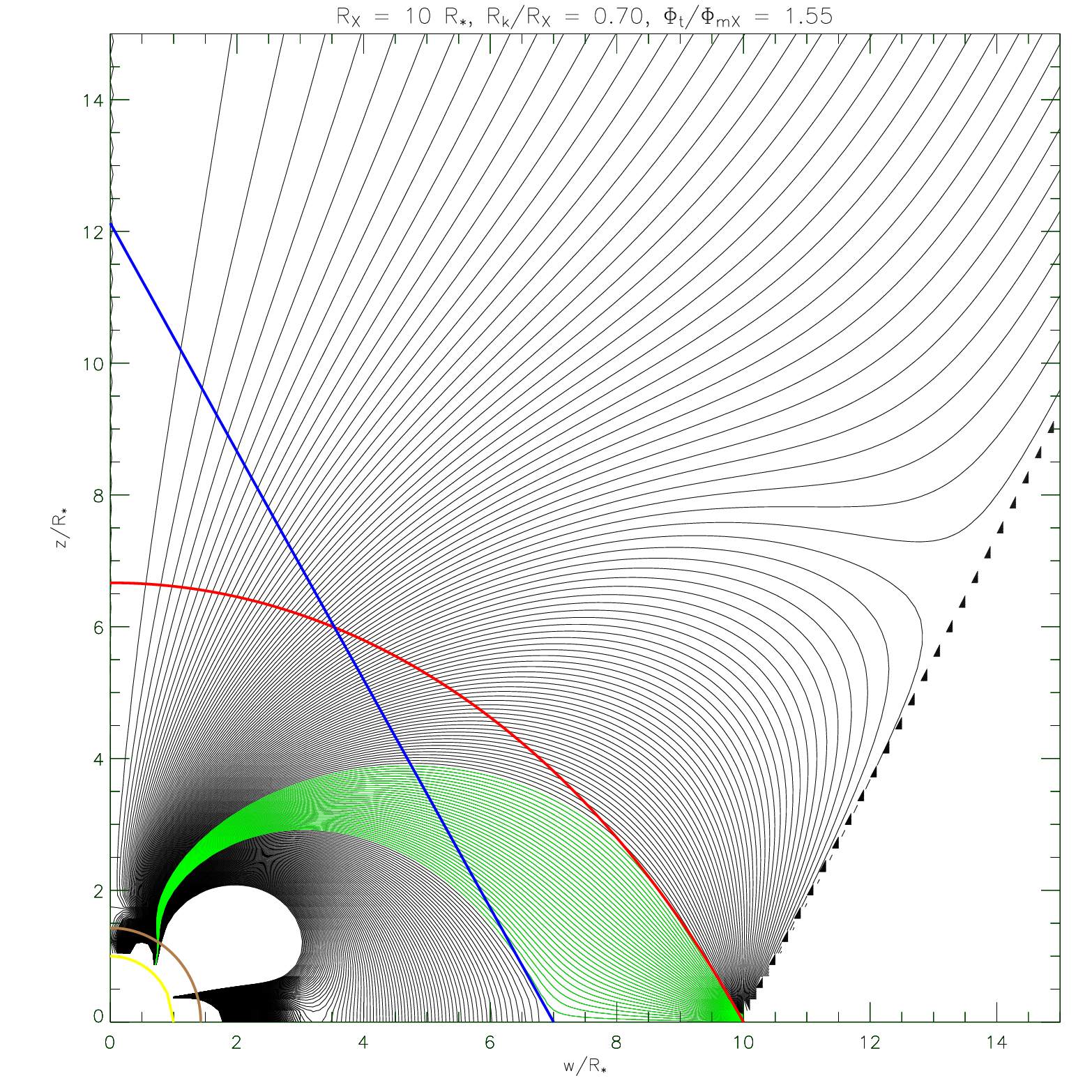}
\figcaption{Final field solution for $\rx$ = 10$\rstar$, with $\rk/\rx$ = 0.70 and $\Phit/\Phimx$ = 1.55.  Contours are lines of constant flux.  Funnel flow field lines in {\it green}, all others in {\it black}.  {\it Dashed line} shows the wind interface.  Equipotential curve shown in {\it red}: the uppermost funnel flow line always lies below this curve, ensuring steady accretion.  Angle of 120$^{\circ}$ to the disk at the kink point $\rk$ shown in {\it blue}: this is the angle at which the last funnel flow line exits the disk at $\rk$.  The {\it yellow curve} represents the stellar surface $R_*$ = 1, on which the original unperturbed field is specified.  The field geometry near $\rx$ closely resembles the dipole solution in Fig. 1, but the funnel is squeezed into a much narrower flow near the stellar surface, producing a very small hot spot (covering fraction 0.065\%) compared to the dipole case.  The {\it brown curve} represents a rescaled stellar surface $R_*$ = 10/7; the $\rx$ = 10$\rstar$ solution shown here for the original surface (in yellow) becomes the $\rx$ = 7$\rstar$ solution for this new surface.  The hot spot covering fraction for the new surface is much larger ($\sim$1\%) than for the original one.  See \S\S 4.1 and 4.2.}
\plotone{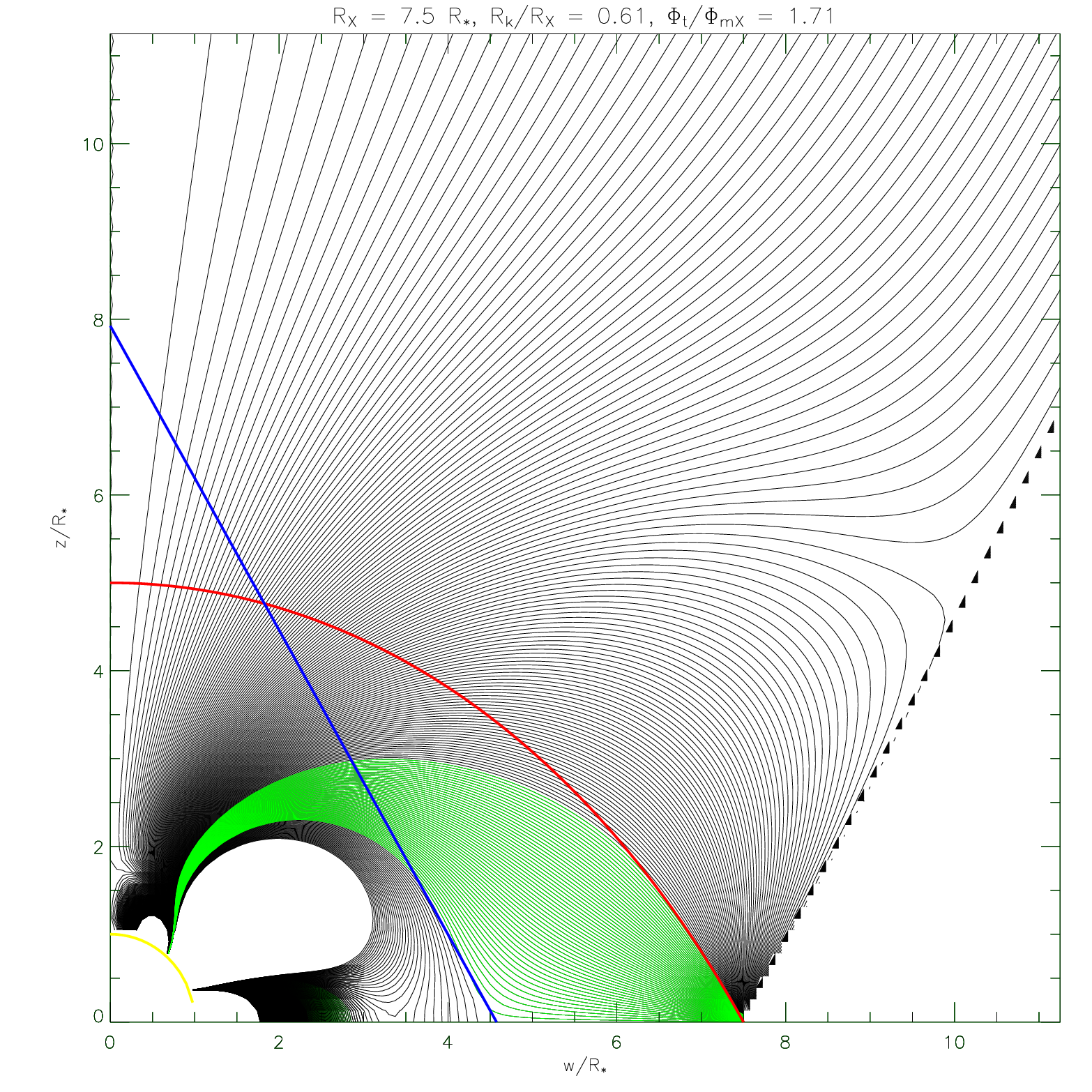}
\figcaption{Final field solution for $\rx$ = 7.5$\rstar$, with $\rk/\rx$ = 0.61 and $\Phit/\Phimx$ = 1.71.  Colors and symbols same as in Fig. 4. See \S\S 4.1 and 4.2. }
\plotone{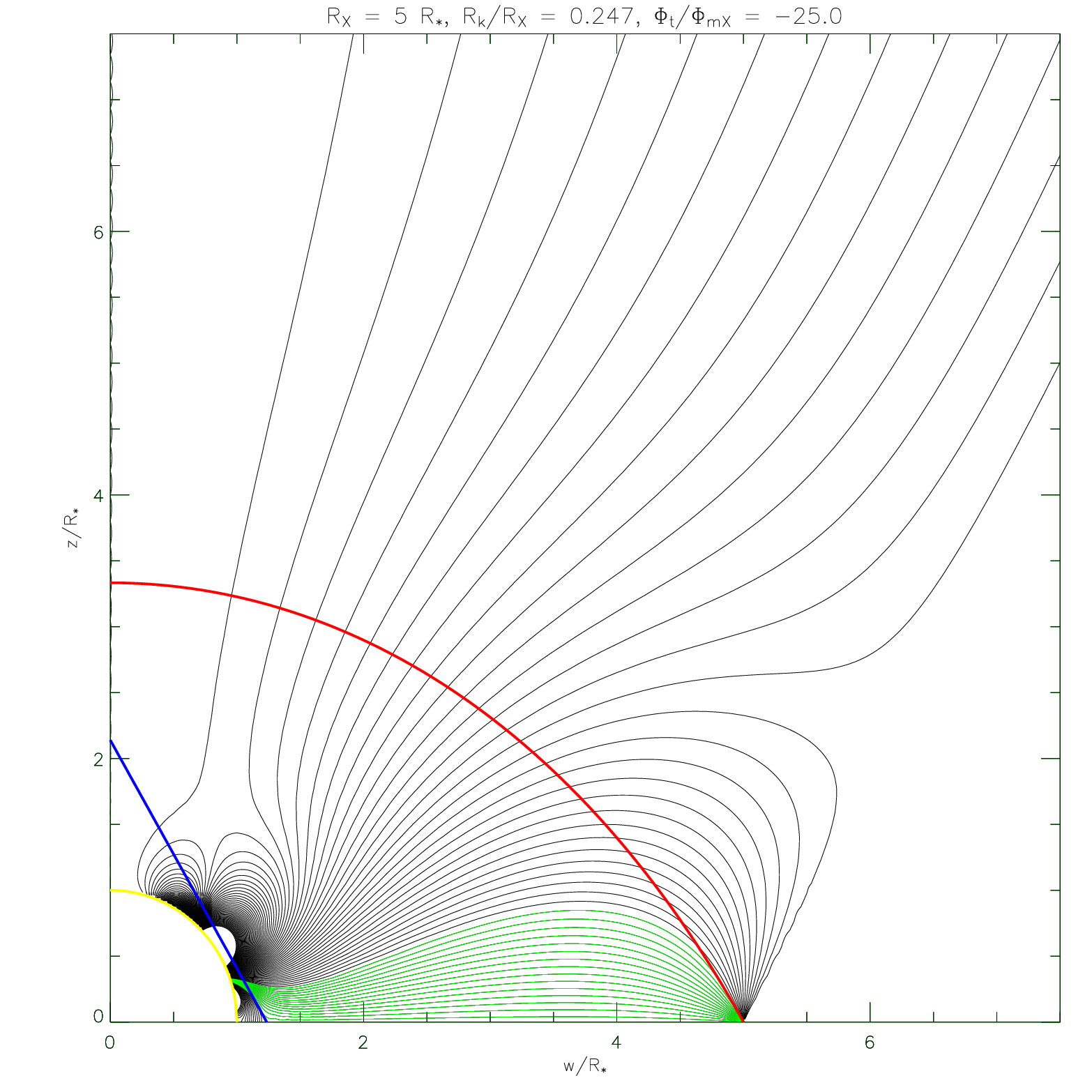}
\figcaption{Final field solution for $\rx$ = 5$\rstar$, with $\rk/\rx$ = 0.247 and $\Phit/\Phimx$ = -25.0.  Colors and symbols same as in Fig. 4.  Note that the hot spot has now jumped to a different location, with the opposite polarity in $B_r$, compared to Fig. 5.  See \S\S 4.1 and 4.2. }
\plotone{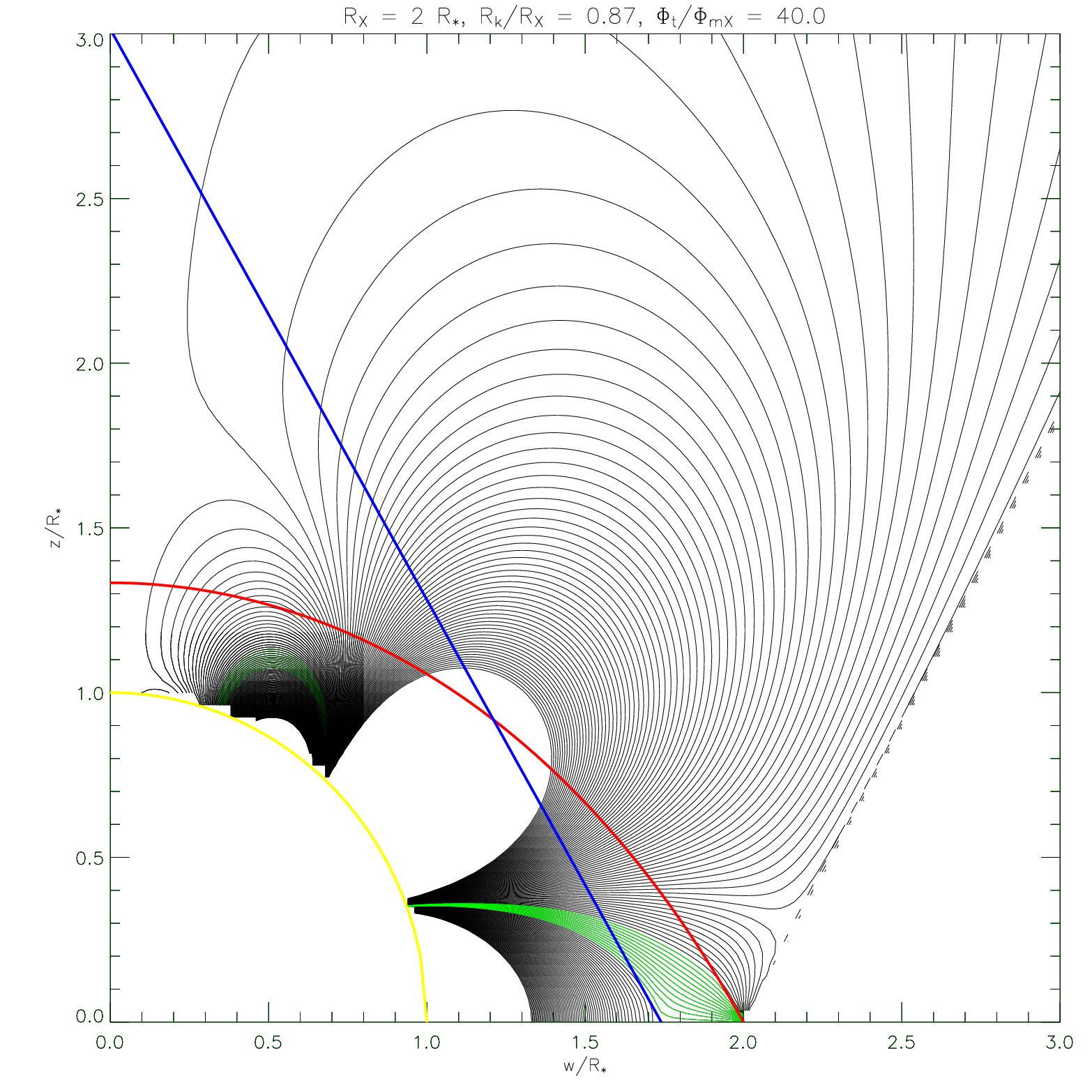}
\figcaption{Final field solution for $\rx$ = 2$\rstar$, with $\rk/\rx$ = 0.87 and $\Phit/\Phimx$ = 40.0.  Colors and symbols same as in Fig. 4.  See \S\S 4.1 and 4.2. }

\begin{deluxetable}{lcc}
\tablecaption{\label{tab1}Parameters of Unperturbed Stellar Multipole Field}
\tablewidth{0pt}
\tablehead{
\colhead{$l$} &
\colhead{$A_l$/$A_1$} \\}   
                            
\startdata

1 & 1\\
3 & +1.64\\
5 & -3.46\\
7 & +0.82\\
9 & +1.45\\
11 & -0.72\\

\enddata

\end{deluxetable}

\end{document}